\documentclass[11pt]{article}

\usepackage[swedish]{babel}
\usepackage[mac]{inputenc}
\usepackage{amssymb, amsfonts, amsmath}
\usepackage{dsfont}
\usepackage{multicol}
\usepackage{graphicx}
\usepackage{indentfirst}
\usepackage{fancybox}
\usepackage{color}
\usepackage[square, comma, numbers, sort&compress]{natbib}
\usepackage{perpage}
\usepackage{adjustbox}

\MakePerPage{footnote} 

\newcommand{\Tr}{\mathrm{Tr}}
\newcommand{\be}{\begin{equation}}
\newcommand{\ee}{\end{equation}}
\newcommand{\bea}{\begin{eqnarray}}
\newcommand{\eea}{\end{eqnarray}}

\newcommand{\wh}{\widehat}

\newcommand{\mc}{\mathcal}
\newcommand{\mr}{\mathrm}
\newcommand{\ti}{\widetilde}
\newcommand{\id}{\mathds{1}}
\newcommand{\ha}{\widehat}
\newcommand{\sgn}{\mathrm{sgn}}
\newcommand{\lie}{\mathfrak}
\newcommand{\vp}{\varphi}

\usepackage{chngcntr}
\begin{document}

\huge
\begin{center}
\textbf{En komprimerad introduktion till \\ 
slingkvantgravitation och spinnskum \\ 
modeller}
\end{center}
\normalsize

\vspace{0.5cm}

\begin{center}
Miklos Långvik\footnote[1]{miklos.langvik@gmail.com}
\end{center}

\vspace{2cm}

\textbf{Abstrakt:} Inom denna reviewarikel ser vi över slingkvantgravitationens och spinnskum modellernas struktur och uppbyggnad. Artikeln 
gör sitt bästa för att förutsätta så lite förkunskaper som möjligt, men en viss kännedom om allmän relativitet i konnektionsvariabler kan vara bra att ha.

\newpage

\tableofcontents

\newpage

\section{Introduktion}

Kvantgravitation är jakten på en teori som förenar den allmänna relativitetsteorin med kvantmekaniken. Denna jakt började redan på 30-talet (se t.ex. \cite{his}, för en kort historik), men trots det fortsätter arbetet av väldigt långt samma frågor än idag. Kvantgravitationsproblemet är svårt eftersom det tangerar många fundamentala begrepp inom vår nutida fysik, begrepp som vi tar för givna men inte vet hur de ter sig på Planck längd ($\sim 10^{-33}$ cm) där kvantgravitationen spelar en roll. Inte minst betydelsen av en tidsvariabel för konstruktionen av en fysikalisk teori och kvantmekanikens basstruktur: linearitet, hermiticitet, etc., men också konstruktionen av en statistisk mekanik innehållande gravitation (se delsektion \ref{sttnw} och definitionen på den termodynamiska tiden \cite{thmtm, thmtm1, thmtm2}) och om det överhuvudtaget går att konstruera en kvant-teori som vid kontinuitetsgränsvärdet återger allmän relativitet? Alla dessa problem gör konstruktionen av en kvantgravitationsteori till ett väldigt svårt problem, men samtidigt ger studierna av frågorna kring kvantgravitationen oss en mycket djupare förståelse av begränsningarna som våra experimentellt verifierade teorier har.
 
Men låt oss börja från början:
Varför tror vi att gravitation skall ha en kvantmekanisk beskrivning på mikronivå? Det bästa svaret på frågan kommer kanske då man betraktar den allmänna relativitetsteorins ekvationer, vilka relaterar rumtidens krökning och därmed geometri till stressenergin inom samma rumtid. Eftersom vi vet att gravitation kopplar till alla 
andra fält som vi experimentellt vet att uppträder kvantiserade på mikronivå, är alltså stressenergin kvantiserad. Men då måste också rumtidens geometri vara det, annars håller inte grundidén bakom Einstein ekvationerna.

Ett annat argument för konstruktionen av en kvantgravitationsteori gäller in\-nandöm\-met av svarta hål. Enligt den allmänna relativitetsteorin finns det en singularitet 
inne i det svarta hålet. Eftersom oändlig stressenergi är nonsens, tror de flesta teoretiker att det finns något sätt att förklara bort denna singularitet. En kvantiserad rumtid skulle kunna sköta problemet eftersom rumtiden i det fallet skulle ha en mista skala och ingen ''oändligt'' liten singularitet torde kunna uppträda.

Det sista argumentet gäller frihetsgraderna som försvinner i ett svart hål. Ifall det som faller in i svarta hål aldrig kan nå oss på nytt i någon form, bryts termodynamikens andra lag.   
Om vi endast förlitar oss på allmän relativitet, kan detta problem inte förklaras. Men ifall vi lyckas konstruera en statistisk mekanik som också innefattar gravitation, borde vi 
via en kvantgravitationsteori kunna förklara vilka dessa frihetsgrader är. Frihetsgrader som det svarta hålet har och som gör att den totala entropin endast kan växa eller hållas konstant. 

Denna reviewartikel kommer att behandla SlingKvantGravitation (SKG, (Loop Quantum Gravity \emph{eng.})), som är ett försök att komplettera Dirac's gamla program att kvantisera gravitation via Hamiltonformalismen (se t.ex. \cite{cbook, thiie, ale} för längre motiveringar och \cite{sahl} för en annan kort introduktion till området), samt spinnskum modeller (eng. \emph{spinfoam models}), vilka är ett försök att tackla problem som uppträder inom SKG (se \cite{alliv, livi, perei} för andra perspektiv på området och 
\cite{zakop} för en definition av EPRL modellen \cite{eprl} utan härledningar.). Detta val av inriktning beror givetvis av skribentens eget 
forskningsintresse, men det kan också motiveras från ett bredare perspektiv. Nämligen är SKG det mest konservativa sättet att se kvantgravitationsproblemet och därför 
borde en kännedom av resultaten i SKG vara av intresse för alla forskare som sysslar med kvantgravitation. Det mest radikala steget för SKG är att kvant-teorin är icke-störningsteoretisk. Detta motiveras med att Einsteins gravitationsteori är bakgrundsfri, d.v.s. det gravitationella fältet existerar inte på en bakgrund, så som ex. det elektromagnetiska fältet propagerar på en Minkowski bakgrund. Det enda som betyder något inom allmän relativitet är relationerna mellan fälten som vi beskriver inom teorin. Om man tar detta som styrande hypotes, måste kvantiseringen av gravitation nödvändigtvis vara icke-störningsteoretisk, eftersom störningar inom gravitation behandlas ovanpå en bakgrund. Trots att begreppet bakgrundsfrihet kräver en del förklaringar, handlar SKG fortfarande om att starta med klassisk gravitation som vi känner den i 
4 dimensioner och sedan kvantisera teorin. Rent beräkningstekniskt blir teorin däremot mer komplicerad p.g.a. sin bakgrundsfrihet, vilken gör användningen av ex. störningsteoretiska approximationer mycket svår. 

SKG är intressant just p.g.a. att den är så konservativ att den ger oss de första insikterna i problemen med kvantiseringen av gravitationen oberoende av vilken
inriktning till kvantgravitationsteori vi är mest intresserade av. Men trots att man kan 
komma väldigt långt med SKG och än är definitivt inte alla kort spelade, har nya modeller, sk. spinnskum modeller börjat uppträda. Dessa är en sorts Feynman stig-integraler inspirerade av SKG. De är konstruerade för att kunna komma åt en totalt Lorentz kovariant konstruktion av kvantgravitation \`a la SKG. Spinnskum modellerna är alltså tänkta att vara en alternativ beskrivning av SKG, men trots att teoriernas fasrum är desamma \cite{eprl, dy1, dy2}, finns det ännu inget fullständigt bevis för deras ekvivalens.

I denna reviewartikel, kommer vi se närmare över strukturen av SKG och spinnskum modellerna. Artikeln delar sig i tre delar, den första handlar om SKG, den andra om SlingKvantKosmologi, SKK (\emph{Loop Quantum Cosmology}) och den sista om spinnskum modellerna. Eftersom SKG kommer att fira 27 års dag i sommar, finns det så mycket att berätta att vi endast kommer att kunna lätt skrapa ytan av området. Hoppeligen kommer denna review trots det att ge en överskådande blick över forskningen inom bakgrundsfri kvantgravitation och för den mera intresserade, en bra grund ifrån vilken man kan läsa in sig bättre på området genom de artiklarna som citeras i denna review. 

I följande sektion går vi igenom SKG:s teoretiska struktur. Vi börjar med att påminna läsaren om Dirac programmet för kvantisering av gravitation i delsektion \ref{skkg}. Sedan 
går vi igenom Hamilton formalismen och 3+1 indelningen användande de speciella variablerna för SKG i delsektion \ref{hamam}. Efter detta konstruerar vi det kinematiska 
Hilbertrummet i \ref{hkkkin} och sedan löser vi 6 av totalt  7 begränsningar (eng. \emph{constraints}) i delsektionerna \ref{spnw} och \ref{diffeo}. Före vi diskuterar problematiken 
med den 7:de begränsningen, tar vi en kort omväg och diskuterar area och volym operatorerna i SKG samt deras egenvärden i delsektion \ref{areavol}. Slutligen diskuterar vi 
problematiken kring den 7:de och sista begränsningen, Hamiltonbegränsningen i \ref{hah}. Efter denna överblick av SKG:s teoretiska struktur, ger vi en kort diskussion om teorins status just nu i delsektion \ref{sttnw}, för att avsluta introduktionen till SKG med en kort om översikt över hur man behandlar svarta hål i denna teori i delsektion \ref{SHSKG}.

\section{SlingKvantGravitation (SKG) \label{skkg}}

SKG är en konsekvens av Dirac programmet för kvantisering av gravitation \cite{dirac} (se \cite{hen} för en mycket belysande, och lång, introduktion till problemen). Därför 
tar vi en kort genomgång av detta program som vi kan indela i fyra steg.
\begin{enumerate}
\item Hitta en representation av fasrumsvariablerna som operatorer i det \emph{kinematiska} Hilbertrummet $\mc{H}_{kin}$. Dessa operatorer satisfierar 
          den ''standardiserade'' kvantiseringen av kommutationsrelationer $\{ , \} \rightarrow -{i\over \hbar}[, ]$. 

\item Promovera teorins begränsningar till själv-adjungerade operatorer i detta kinematiska Hilbertrum.  

\item Karakterisera lösningsrummet (eng. \emph{space of solutions}) till begränsningsoperatorerna och konstruera inre produkten till 
det slutliga, \emph{fysikaliska} Hilbertrummet $\mc{H}_{phys}$. Denna inre produkt definierar den fysikaliska sannolikheten i teorin. 

\item Hitta en fullständig mängd gauge invarianta observabler, som kommuterar med begränsningsoperatorerna. Dessa representerar de frågor som 
kan besvaras med teorin. 
\end{enumerate}
För att komma åt att besvara dessa steg startar man i SKG från Holst aktionen \eqref{holst} som man konstruerar en Hamiltonformalism för genom ett speciellt 
val av variabler. Detta definierar begränsningarna som man sedan kvantiserar. I SKG har vi en väldefinierad kvantisering av alla dess 7 begränsningar, men inte på 
$\mc{H}_{kin}$ utan på diffeomorfisminvarianta kombinationer av tillstånd i $\mc{H}_{kin}$. Detta för att den Hamiltonska begränsningen, den som står i relation till koordinattiden, är mycket tvetydig och därför svår att kvantisera. Därför kan man påstå att vi i SKG för tillfället står någonstans mellan steg 2 och 3 i Dirac programmet. I det följande beskriver vi denna utveckling i mera exakta steg.  

\subsection{Klassisk Hamilton formalism \label{hamam}}

I SKG är startpunkten Holst aktionen \cite{holle} (i enheter $\hbar = c = 1$) 
\begin{align}
S[e, \omega] = {1\over 8\pi l_p^2} \int \epsilon^{IJKL} e_I\wedge e_J\wedge F_{KL}(\omega) + {1\over \gamma}e^I\wedge e^J\wedge F_{IJ}(\omega), \label{holst}
\end{align}
där $\omega$ är en SL(2, $\mathbb{C}$) konnektion, $F_{IJ}$ dess krökning (eng. \emph{curvature}), $e^I_{\mu}$ en tetrad, $\gamma$ Barbero-Immirzi parametern \cite{immi} och  $l_p$ Planck längden. Holst aktionen är den vanliga Palatini aktionen + Holst termen, den med $\gamma$. Denna Holst-term ändrar inte på rörelse-ekvationerna för gravitation och nämns därför oftast som en topologisk term. Detta är inte helt korrekt, eftersom Holst-termen har geometriska konsekvenser när man kopplar gravitation till fermioner \cite{fermhol}, men vi håller oss till renodlad gravitation i denna korta presentation och bryr oss därför inte mera om detta. 

I nästa steg splittrar vi Holst aktionen i 3+1 dimensioner enligt ADM metoden \cite{adm}. För att göra detta antar vi att mångfaldet $\mathcal{M}$ vi konstruerar teorin på, är globalt hyperboliskt. Detta ger oss möjligheten att splittra $\mathcal{M}$ i $\mathcal{M} = \mathbb{R} \times \Sigma$, där $\Sigma$ rums-mångfaldet har en fixerad topologi och $\mathbb{R}$ innehar koordinattiden. Denna 
process fixerar topologin i SKG, men visar också vikten av $\gamma$-parametern. I den vanliga ADM konstruktionen 
av Hamiltonsk gravitation där $\gamma = 0$ är de kanoniska variablerna den 3-dimensionella rums-metriken och den extrinsiska krökningen. I SKG med reella variabler 
\cite{barb} är de däremot en SU(2) konnektion $A^i_a$ och en kanonisk rörelsemängd $E^b_j = {1\over 2}\epsilon_{ijk}\epsilon^{abc}e^j_be^k_c$, där $e^j_b$ är en triad, som satisfierar
\be
\{A_a^i(x), E_j^b(y)\} = 8\pi l_p^2\gamma \delta_a^b\delta_j^i\delta(x, y), \label{pois}
\ee
där $\{, \}$ är en Poison-kommutator och de andra Poison kommutatorerna av de kanoniska variablerna är 0. Det är klart att utan $\gamma$-parametern, skulle variablerna Poison-kommutera. Den geometriska tolkningen av dessa variabler är relativt enkel. Eftersom $qq^{ab} = E^a_iE^b_j\delta^{ij}$, där $q^{ab}$ är den 3 dimensionella intrinsiska krökningen på $\Sigma$ och $q$ dess determinant, innehåller $E^a_i$ all geometrisk information på $\Sigma$. $A^a_i$ i sin tur är en kombination av spinn-konnektionen $\Gamma^a_i$ och den extrinsiska krökningen $K^a_i$ på $\Sigma$ enligt $A^a_i = \Gamma^a_i + \gamma K^a_i$. I den gamla formuleringen av SKG med sk. Ashtekar variabler \cite{ashold} väljer man $\gamma = \pm i$ och konnektionen är en självdual SL($2, \mathbb{C}$) konnektion, men nuförtiden är det vanligare att välja $\gamma \in \mathbb{R}$. Vi kommer snart att se vad detta val innebär, men först lite mer information om den Hamiltonska konstruktionen.

Då vi genomfört 3+1 splittringen för variablerna $A_a^i$ och $E_j^b$ kommer vi att ha en Hamiltonfunktion, lineär i första klass begränsningar, så som 
en bakgrundsfri teori kräver, eftersom gravitation är en gaugeteori och all gauge måste vara av första klass.  
Dessa begränsningar genererar gaugetransformationer inom fasrummet. De är
\begin{align}
\mathcal{G}_i &= D_aE_i^a = \partial_aE^a_i + \epsilon_{ij}^{\, \, \, \, \, k}A^j_aE^a_k, \label{const1} \\
V_a &= E^b_i F^i_{ab}, \label{const2} \\
H &= {E^a_iE^b_j\over \sqrt{\det E}} \big({1\over 2}\epsilon^{ij}_{\, \, \, \, \, k} F^k_{ab} - (1 + \gamma^2)K^i_{[a}K^j_{b]} \big), \label{const3}
\end{align}
där $F^i_{ab}$ är krökningen av SU(2) konnektionen, $D_a$ dess kovarianta derivata och $K^i_a$ den extrinsiska krökningen på $\Sigma$.
Den första begränsningen \eqref{const1} genererar SU(2) transformationer och kallas därför för Gauss-begränsningen eftersom den är analog med
begränsningen i Gauss lag inom elektrodynamik. Den andra begränsningen \eqref{const2} genererar transformationer i fasrummet från 
rum-diffeomorfismer på $\Sigma$ och kallas för 
vektorbegränsningen. Den tredje \eqref{const3} genererar transformationer i fasrummet från deformationer av $\Sigma$ i en tidslik riktning\footnote[1]{Det bör nämnas att då SU(2) konnektionen valdes, fastställdes gauge-måttet på konnektionen partiellt, så att vi från $\omega$ i SL(2, $\mathbb{C}$) hamnade med $A^a_i$ i SU(2). Detta resulterar i den tidslika riktningen som nämndes. Det partiella gauge-måttet som valts är en direkt konsekvens av att vi närmar oss gravitationsproblemet via Hamiltonformalismen.} av rumtiden och kallas för Hamiltonbegränsningen. Det är den som ger flest
problem inom SKG eftersom den är svår att lösa exakt. Därför skulle de gamla själv-duala Ashtekar variablerna med $\gamma = \pm i$ vara lockande. Men i detta fall antar aktionen \eqref{holst} komplexa värden. Då måste man addera sk. realitets villkor (eng. \emph{reality conditions}) till teorin (se t.ex. \cite{reall}) och dessa har visat sig vara ännu jobbigare 
än \eqref{const3}. Därför har man börjat hålla fast vid en reell $\gamma$. Trots nödvändigheten av Barbero-Immirzi parametern för teorins matematiska konstruktion, finns det än så länge ingen välförståd fysikalisk orsak bakom den. Detta är något man hoppas hitta inom studien av denna teori. 

\subsection{Hilbertrummet $\mathcal{H}_{kin}$ \label{hkkkin}}

För att bygga upp SKG som en kvantteori, måste vi definiera det fysikaliska Hilbertrummet för teorin. Detta sker genom ett antal steg. Först definierar man ett Hilbertrum 
som kallas för det kinematiska Hilbertrummet $\mathcal{H}_{kin}$, vilket är Hilbertrummet som vi löser begränsningarna \eqref{const1}-\eqref{const3} på. Detta sker en begränsning i taget och man brukar nämna dem enligt följande: $\mathcal{H}^\mathcal{G}_{kin}$ är delmängden av $\mathcal{H}_{kin}$ som löser begränsningen \eqref{const1}, 
$\mathcal{H}^{\mathrm{Diff}}_{kin}$ är delmängden av $\mathcal{H}_{kin}$ som löser både \eqref{const1} och \eqref{const2} och slutligen är Hilbertrummet $H_{phys}$, det fysikaliska Hilbertrummet för teorin som är delmängden av $\mathcal{H}_{kin}$ som löser alla begränsningar. Att hitta dessa delmängder av $\mathcal{H}_{kin}$ är inte alls så enkelt att göra i praktiken som det är att säga, men för att se vad vi kan göra och 
var vi får problem börjar vi med att definiera $\mathcal{H}_{kin}$. 

För att göra detta måste vi börja med att definiera algebran av sk. kinematiska observabler 
som man använder sig av för att konstruera $\mathcal{H}_{kin}$ inom SKG (Grundidéerna till denna konstruktion fanns redan i \cite{gamb, rovsmo}, men de har fått sin
slutliga struktur i \cite{ashish, ashlew, lew, almea, jere}). Denna algebra kallas för holonomi-flödes algebran (eng. \emph{holonomy-flux algebra}) och dess grundvariabler består av holonomin för SU(2) konnektionen
\be
h_e[A] = P\int_e \exp(A), \label{holo}
\ee
där $e$ är stigen man integrerar linje integralen längs, P är stig ordnings-symbolen (eng. \emph{path ordering symbol}), samt flödet av $E^a_i$ genom en yta $S$ 
\begin{align}
E[S,f] = \int_S \star E_if^i = \int_S d\sigma^1d\sigma^2{\partial x^a\over \partial\sigma^1}{\partial x^b\over \partial\sigma^2} f^i {\delta \over \delta A^i_c} \epsilon_{abc} \label{flux}
\end{align}
där $f^i$ är en funktion som antar värden i $\mathfrak{su}$(2) och $\star E_i = E^a_i\epsilon_{abc}dx^b\wedge dx^c$. Vi ser att flödet genom en yta \eqref{flux} kommer 
att agera derivata i vår algebra. Detta kan ses som en konsekvens av att de nya variablerna antar likadana Poison kommutatorer som i \eqref{pois}. Deras fördel är 
att man kan använda dem för att konstruera tillstånd i $\mathcal{H}_{kin}$ m.h.a. vilka lösningarna till Gauss- \eqref{const1} och vektorbegränsningarna \eqref{const2} lätt hittas. 

\begin{figure}[h!]
\begin{center}
\includegraphics[width=8cm]{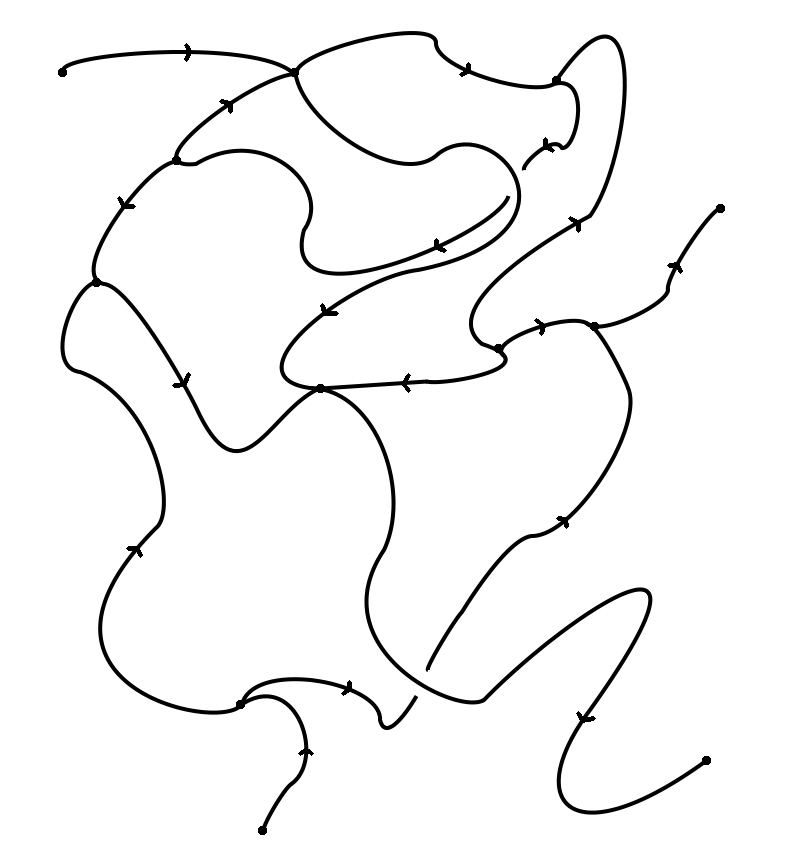}
\end{center}
\caption{Vi ser en graf med 15 stigar med sina respektive riktningar (pilarna) och 12 noder.}
\label{snran1}
\end{figure}

Denna algebra kallas för Holonomi-flödes algebran och består av algebran som alla cylindriska funktioner på $\Sigma$ skapar. En cylindrisk funktion definieras i detta fall som
\be
\psi_{f, \gamma}[A] = f(h_{\alpha_1}[A], h_{\alpha_2}[A],..., h_{\alpha_{N_e}}[A]), \label{cyl}
\ee 
där $f\, :\, \mathrm{SU(2)}^{N_e} \rightarrow \mathbb{C}$ och $\gamma$ är en graf med $N_e$ stycken stigar. En graf definieras som en ändlig ansamling stigar med en riktning
från stigens början till dess slut på $\Sigma$ som endast möts i sina ändor, om de möts överhuvudtaget (se Figur \ref{snran1}). 
Detta betyder att algebran av de cylindriska funktionerna Cyl, matematiskt kan definieras som
\be
\mathrm{Cyl} = \cup_\gamma \mathrm{Cyl}_\gamma, 
\ee 
där Cyl$_{\gamma}$ representerar en cylindrisk funktion för grafen $\gamma$ och $\cup_\gamma$ är unionen av alla grafer på $\Sigma$. Flödesvariabeln \eqref{flux}
agerar derivata $X_{S, f}$ i denna algebra \cite{deriv} enligt följande definition
\be
X_{S, f}\psi_{g, \gamma} = {1\over8\pi\gamma l_p^2}\{\psi_{g, \gamma}, E_{S, f}\}.
\ee
För att fullfölja vår konstruktion av det kinematiska Hilbertrummet $\mathcal{H}_{kin}$, konstruerar vi representationen av dess tillstånd enligt
\be
\mu_{AL}(\psi_{f, \gamma}) = \int\prod_{e\, \subset\, \gamma}dh_ef(h_{e_1},h_{e_2},...,h_{e_{N_e}}), 
\ee
där $h_e \in$ SU(2) och $dh_e$ är det normaliserade Haar måttet för SU(2). Detta tillstånd kallas för Ashtekar-Lewandowski måttet \cite{almea}. Av denna definition följer direkt $\mu_{AL}(1) = 1$ samt $\mu_{AL}(\overline{\psi_{f, \gamma}}\psi_{f, \gamma}) \geq 0$. Vi definierar därmed inre produkten för de cylindriska funktionerna som
\begin{align}
\langle \psi_{f, \gamma}| \psi_{g, {\gamma'}} \rangle = \mu_{AL}(\overline{\psi_{f, \gamma}}\psi_{g, {\gamma'}}) = \int\prod_{e\, \subset\, \Gamma_{\gamma\gamma'}}
dh_e\overline{f(h_{e_1},...,h_{e_{N_e}})}g(h_{e_1},...,h_{e_{N_e}}), \label{inprod} 
\end{align}
där $\Gamma_{\gamma\gamma'}$ är en graf för vilken både $\gamma \subset \Gamma_{\gamma\gamma'}$ och  $\gamma \subset \Gamma_{\gamma\gamma'}$ gäller. 
Vi definierar slutligen $\mathcal{H}_{kin}$ som Cauchy kompletteringen av rummet för de cylindriska funktionerna i Ashtekar-Lewandowski måttet. De cylindriska funktionerna $\psi_{f, \gamma}[A] = \langle A|\psi_{\gamma, f}\rangle = f(h_{e_1},...,h_{e_{N_e}})$ blir på så sätt tillstånd i Hilbertrummet $\mathcal{H}_{kin}$. 
 
\subsection{Spinn-nätverk \label{spnw}}

Efter att vi byggt upp och definierat $\mathcal{H}_{kin}$ kan vi nu ange en ortonormal bas för $\mathcal{H}^\mathcal{G}_{kin}$, delmängden av $\mathcal{H}_{kin}$ som 
löser begränsningen \eqref{const1}. Denna består av spinn-nätverk \cite{reis, roso, baez} (se också \cite{smol} för en diskussion om hur spinn-nätverk kan användas inom topologisk kvantfältteori och gaugeteori). För att introducera spinn-nätverken är det bäst att påminna sig om holonomins \eqref{holo} följande egenskaper
\begin{align}
h_e[A] &= h_{e_1}h_{e_2} \label{divholo} \\
h_{e^{-1}}[A] &= h^{-1}_e[A] \label{invholo} \\
h'_e[A] &= g(x(0))h_e[A]g^{-1}[x(1)]. \label{gaugeholo}
\end{align} 
\eqref{divholo} demonstrerar hur holonomin av en stig $e$ med riktning (eng. \emph{orientation}) delas i två delar  $e_1$ och $e_2$, där slutet av stigen $e_1$ limmas fast i början av stigen $e_2$. \eqref{invholo} visar hur holonomin beter sig för en stig med inverterad riktning och i \eqref{gaugeholo} ser vi gaugetransformationen av holonomin för en stig 
$e$ där är $x(0)$ början på stigen och $x(1)$ dess slut. 

\begin{figure}[h!]
\begin{center}
\includegraphics[width=6cm]{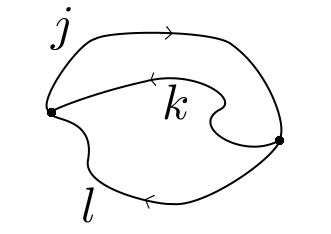}
\end{center}
\caption{Figuren visar ett spinn-nätverk med två noder och tre stigar med spinn $j,k$ och $l$. Pilarna på stigarna indikerar i vilken nod stigen startar och i vilken 
nod den slutar.}
\label{snwrk}
\end{figure}

Id\'een bakom valet av holonomin som en av grundvariablerna i teorin är helt enkelt dess transformationsegenskaper under begränsningarna \eqref{const1} och 
\eqref{const2}\footnote[1]{Se sektion \ref{diffeo}}. T.ex. kan vi bilda Wilson slingan\footnote[2]{härav namnet SlingKvantGravitation}
\be
W_\sigma[A] = \Tr[h_\sigma[A]], \label{wil}
\ee  
för slingan $\sigma$, där man med slinga menar en sluten stig. Denna kvantitet är, p.g.a. $\Tr$-operationen, invariant under gaugetransformationerna \eqref{gaugeholo}. 
Det bästa är att $W_\sigma[A]$ är ett element i Cyl$_{\sigma}$, d.v.s. vi har kommit åt en del av Cyl som också löser \eqref{const1}. För att generalisera denna id\'e märker vi att
vi p.g.a. egenskapen \eqref{divholo}, kan bl.a. dela på slingan i två stigar $e_1$ och $e_2$ som sitter fastbundna i varandras ändor. Då får vi Wilson slingan 
$W_\sigma[A] = \Tr[h_{e_1}[A]h_{e_2}[A]]$ som är en del av Cyl$_{e'}$, där Cyl$_{e'}$ består av de cylindriska funktionerna för stigarna $e_1$ och $e_2$ och deras kombinationer. Ett sätt att ytterligare generalisera detta är att välja en irreducerbar SU(2)-representationsmatris $D$ och representera holonomin 
med denna i \eqref{wil}, vilket ger oss
\be
W^D_\sigma[A] = \Tr[ D(h_\sigma[A])]. \label{wil_int}
\ee
Denna slinga är också gaugeinvariant. Om vi anger unitära irreducerbara representationsmatriser av spinn $j$ för SU(2) i formen $D^j_{mm'}$ för 
$-j \leq m,m' \leq j$, är den cylindriska funktionen
\be
W^j_\sigma[A] = \Tr[ D^j(h_\sigma[A])], \label{simpspin}
\ee
det enklaste exemplet på ett spinn-nätverk. För att konstruera ett större spinn-nätverk med flere stigar, måste vi ännu införa ett gaugeinvariant sätt att multiplicera
representationsmatriserna i spinn-$j$ representationen. Detta går enkelt genom sk. sammanflätare (eng. \emph{intertwiners}), invarianta tensorer i tensorprodukten av 
SU(2) representationsmatriserna för de stigar vars ändor slutar i samma nod. Detta är bäst att illustrera genom ett exempel. Vi väljer en graf med tre stigar (se Figur \ref{snwrk}), 
vilka bär på representationsmatriserna i spinn $j,k$ och $l$ representationen enligt Figur \ref{snwrk}. 
Då kan vi bilda spinn-nätverksfunktionen
\begin{align}
\Theta^{j,k,l}_{e_1 \cup e_2 \cup e_3} = D^j_{m_1n_1}(h_{e_1}[A])D^k_{m_2n_2}(h_{e_2}[A])D^l_{m_3n_3} (h_{e_3}[A]) \mathtt{i}^{m_1m_2m_3}\mathtt{i}^{n_1n_2n_3},   
\end{align}
där $\mathtt{i}^{m_1m_2m_3}$ är en sammanflätare i tensorprodukten $j \otimes k \otimes l$. M.a.o. kan vi bilda cylindriska funktioner 
som spinn-nätverks-funktioner
\be
s_{\gamma, \{j_e\}, \{\mathtt{i}_n\}}[A] = \underset{n \subset \gamma}{\otimes}\mathtt{i}_n\underset{e \subset \gamma}{\otimes}\overset{j_e}D(h_e[A]),
\ee
där summeringen av de invarianta tensorerna $\mathtt{i}$ och SU(2) representationerna $D$ lämnats implicit och notationen $n \subset \gamma$ och $e \subset \gamma$ står 
för respektive en nod i grafen $\gamma$ och en stig i grafen $\gamma$. Dessa spinn-nätverksfunktioner är gaugeinvarianta under SU(2) och därför lösningar på \eqref{const1}. Utöver detta bildar de en ortonormal bas i $\mathcal{H}^\mathcal{G}_{kin}$, vilket kan visas via Peter-Weyl teoremet som är en sorts generalisering av ''Fourier''-analys 
på $S^1$ till kompakta Lie-grupper. Ett kort bevis på att spinn-nätverken verkligen anger en ortonormal bas kan hittas i \cite{ale}. 

Vårt slutresultat består av att vi hittat en delmängd av de cylindriska funktionerna och en grafisk representation av dessa som löser \eqref{const1} och bildar en 
ortonormal bas i $\mathcal{H}^{\mathcal{G}}_{kin}$ utan att någonsin behöva lösa \eqref{const1} separat för sig: spinn-nätverk.

\subsection{Diffeomorfismerna på $\Sigma$ \label{diffeo}} 
 
Holonomi-flödes variablerna är inte endast praktiska för att hitta $\mathcal{H}^{\mathcal{G}}_{kin}$, de är också väldigt användbara 
för konstruktionen av $\mathcal{H}^{\mathrm{Diff}}_{kin}$, Hilbertrummet som innehåller lösningarna till både \eqref{const1} och \eqref{const2}. Detta
följer från holonomins transformationsegenskaper under diffeomorfismer på $\Sigma$
\begin{align}
h_e[\phi^*A] = h_{\phi^{-1}(e)}[A] \label{diffholo} 
\end{align} 
där $\phi^*A$ är aktionen av diffeomorfismen $\phi \in \mathrm{Diff}(\Sigma)$ på konnektionen $A$. P.g.a. \eqref{diffholo} kan vi 
konstruera en unitär operator $\ha{\mathcal{U}}_{\mathrm{Diff}}[\phi]$, där $\phi$ är en diffeomorfism på $\Sigma$, som via sin verkan på Cylindriska funktioner
i $\mathcal{H}_{kin}$
\be
\ha{\mathcal{U}}_{\mathrm{Diff}}[\phi]\psi_{f,\gamma}[A] = \psi_{f,\phi^{-1}\gamma}[A], \label{difftrans}
\ee 
ger oss en cylindrisk funktion som är invariant under rum-diffeomorfismerna på $\Sigma$. Det finns dock två komplikationer: i) Funktionen i \eqref{difftrans}
befinner sig i det topologiska dual rummet Cyl$^*$ och ii) det
existerar inte någon väl-definierad själv-adjungerad generator av infinitesimala diffeomorfismer för den unitära operatorn \eqref{difftrans} eftersom \eqref{difftrans}
inte är svagt kontinuerlig topologiskt. 

Problem ii) sköter vi lätt eftersom operatorn \eqref{difftrans} är allt vi behöver för diffeomorfism invarianta tillstånd och vi kan därför 
ersätta begränsningen \eqref{const2} med 
\be
\ha{\mathcal{U}}_{\mathrm{Diff}}[\phi]\psi_{f,\gamma}[A]  = \psi_{f,\gamma}[A], \label{diffI} 
\ee
som solklart är en ekvivalent begränsning. Problem i) som beror av att gauge-omlopps\-banorna (eng. \emph{gauge-orbits}) 
för diffeomorfismerna inte är kompakta är lite mer involverat att lösa. Id\'en är att använda distributionella tillstånd \cite{almea} och den inre produkten \eqref{inprod} 
för att projicera ut den diffeomorfisminvarianta informationen för tillståndet. Vi bygger det distributionella tillståndet enligt
\be
([\psi_{f, \gamma}]| = \sum_{\phi \in \mathrm{Diff}(\Sigma)} \langle \psi_{f, \gamma}|\ha{\mathcal{U}}_{\mathrm{Diff}}[\phi], \label{diffdist}
\ee
där summan går över alla diffeomorfismer $\phi$ på $\Sigma$ och $[\psi_{f, \gamma}]$ betyder att det distributionella tillståndet endast beror av 
ekvivalensklasserna för $\phi$ under diffeomorfismer. Med dessa distributionella tillstånd, kan vi definiera en inre produkt 
\be
\langle [\psi_{f, \gamma}]|[\psi_{g, \gamma'}] \rangle_{\mathrm{Diff}} = ([\psi_{f, \gamma}]|\psi_{g, \gamma'} \rangle, \label{diffprod}
\ee 
för Hilbertrummet $\mathcal{H}^{\mathrm{Diff}}_{kin}$ \cite{skno}. Den inre produkten \eqref{diffprod} är förstås diffeomorfisminvariant, vilket följer direkt från \eqref{diffI}. 

Vi har nu byggt upp ett Hilbertrum  $\mathcal{H}^{\mathrm{Diff}}_{kin}$ som består endast av lösningar av både \eqref{const1} och \eqref{diffI}\footnote[1]{Dessa 
tillstånd är relaterade till sk. s-knutar \cite{skno} (eng. \emph{s-knots})}. Dessutom 
visar det sig att representationen av de cylindriska funktionerna i $\mathcal{H}^{\mathrm{Diff}}_{kin}$, som vi valt, är unik \cite{lost, lost1}. Mera precist är representationen 
unik för vilken som helst diffeomorfisminvariant teori som använder en kompakt konnektion som konfigurationsvariabel och byggs upp via holonomi-flödes variabler.  
 
\subsection{Kvantgeometri \label{areavol}} 
 
Innan vi ser på svårigheterna med Hamiltonbegränsningen \eqref{const3}, visar vi här hur SKG leder till en geometri som är kvantiserad. 

\subsubsection{Area}

Vi börjar genom att visa hur arean är kvantiserad inom SKG \cite{almea, smop, rovsm}. Arean i SKG definieras via ett gränsvärde av en Riemann summa
\be
A_S = \underset{N \rightarrow \infty}{\lim} A^N_S, \label{limA}
\ee 
där $S \subset \Sigma$ är en yta, som är delad i 2-celler. Riemann summan anges av
\be
A^N_S = \sum_{I=1}^N\sqrt{E_i(S_I)E^i(S_I)},
\ee 
där $N$ är antalet 2-celler och $E_i(S_I)$ är flödet genom den I:tte cellen. Gränsvärdet \eqref{limA} definierar arean på 
ytan $S$ i klassisk Riemannsk geometri. I vårt fall räcker det med att kvantisera flödet $E_i(S_I)$, vilket vi redan gjort i \eqref{flux}. Nu kan 
vi beräkna flödet genom en yta som korsas endast en gång av en stig. Först beräknar vi följande derivata av holonomin
\begin{align}
{\delta\over \delta{A^i_e}}h_e[A] = {\delta\over \delta{A^i_e}}(P\exp\int ds \dot{x}^d(s)A^k_d \tau_k) = \int ds \dot{x}^e(s)\delta^3(x(s) - x)h_{e_1}[A]\tau_ih_{e_2}[A], \label{derholo}
\end{align}
där $\tau_i$ är en generator av SU(2). Vi ser att derivatan splittrar holonomin i två delar på båda sidor om punkten där en agerar. Eftersom
denna derivata är del av \eqref{flux}, kan vi beräkna flödet genom en yta som korsas endast en gång av en stig som 
\begin{align}
E_i(S_I)E^i(S_I)D^j_{mn}(h_e[A]) = (8\pi\gamma l_p^2)^2 j(j+1)D^j_{mn}(h_e[A]), \notag
\end{align}   
där $j(j+1)$ faktorn följer av att $\tau^i\tau_i = j(j+1)$, m.a.o. Casimiroperatorn för SU(2). Detta implicerar i sin tur direkt att
areaoperatorn i SKG anges av
\be
\ha{A}_S |\psi_{f, \gamma}\rangle = 8\pi \gamma l_p^2 \sum_p \sqrt{j_p(j_p + 1)}|\psi_{f, \gamma}\rangle, \label{kvar}
\ee
där vi har en summa över alla stigar som punkterar ytan $S$ som vi beräknar arean på. Resultatet \eqref{kvar} visar att vi har en kvantiserad area i SKG. 

Innan vi fortsätter till volymen finns det några saker att nämna om arean. För det första torde det vara klart att vi måste regularisera areaoperatorn på något vis om vi skall 
kunna gå från steg \eqref{limA} till steg \eqref{kvar}. I SKG använder sig av en lite annorlunda mening för ordet regularisering än i kvantfältteori. Med regularisering menas
i SKG att man skriver om en integral av en klassisk kvantitet som en Riemannsk summa av termer som man efter detta steg promoverar till operatorer. Aktionen 
av denna operator på en fixerad spinn-nätverks funktion återger, då summan är oändlig, den klassiska kvantiteten, integralen. Om denna kvantitet kan förfinas så, att den 
i något skede inte längre beror av storleken på ''stegen'' i Riemann summan, säger vi att den är oberoende av regularisering. En viktig konsekvens av denna procedur 
är att UltraVioletta (UV) singulariteter inte kan existera i SKG.

Då vi regulariserar arean som en Riemannsk summa av 2-celler enligt \eqref{limA}, måste vi göra det så, att de kan minskas till så små 2-celler att endast en stig per 2-cell 
korsar ytan. Annars kan vi hamna i situationen där en och samma stig korsar ytan flere
gånger än endast en gång, eller i en situation där det finns flere stigar än en som korsar ytan. I dessa situationer kommer inte \eqref{derholo}
att vara av lika enkel form och slutresultatet \eqref{kvar} förstörs. 

Den andra saken som är av mera fysikalisk betydelse är att 
trots att \eqref{kvar} visar att arean säkert är kvantiserad inom denna teori, är detta inte nödvändigtvis det slutliga 
areaspektret eftersom vi inte ännu löst Hamiltonbegränsningen \eqref{const3}. Trots detta, finns det redan nu fall då vi vet att arean säkert är kvantiserad
enligt \eqref{kvar}. Detta händer då man betraktar svarta hål enligt SKG filosofin (se sektion \ref{SHSKG}). 

\subsubsection{Volym} 

Volymoperatorn \cite{smop, rovsm, loll, ask} är inte lika väl-definierad som arean och den beror också på hur man väljer att regularisera den. För tillfället 
finns det åtminstone två olika regulariseringar av volymoperatorn som leder till mycket liknande men trots det olika volymoperatorer \cite{rovsm, ask}. Dessutom
är volymspektret än så länge okänt, eftersom det visat sig vara väldigt svårt att beräkna rent beräkningstekniskt, det enda som
är säkert är att det är diskret. 

För att hitta volymoperatorn, startar vi ifrån dem klassiska definitionen för en 3-dimensionell volym $B \subset \Sigma$
\be
V_B = \int_B\sqrt{q}d^3x,
\ee
där $q$ är determinanten av 3-metriken $q^{ij} = e^a_ie^b_j\delta^{ij} $ som definierar $E^i_a = {1\over 2}\epsilon_{ijk}\epsilon^{abc}e^j_be^k_c$ 
(se paragrafen före \eqref{pois}). M.h.a. dessa identiteter kan vi skriva om $V_B$ som
\be
V_B = \int_B\sqrt{\Big|{1\over 3!}\epsilon_{abc} \epsilon^{ijk}E^a_iE^b_jE^c_k \Big|}d^3x.
\ee   
Nu kan vi använda oss av samma teknik som vi använde för areaoperatorn för att definiera volymoperatorn för SKG. Vi delar 
in volymen vi beräknar i små 3-celler så att vi kan bilda en Riemannsk summa av dem. I varje 3-cell anger vi 3 ytor som korsar varandra, vilka
behövs för att beräkna aktionen av flödesoperatorerna $E^a_i$. Volymoperatorn kommer således att ge en kontribution då en
en nod av spinn-nätverket sammanfaller med korsningen av de tre ytorna. Vi kan mycket väl tänka oss en situation där vi endast har 
en stig utan noder som korsar ytorna i tre olika punkter i 3-cellen, men eftersom sådana situationer kommer att kunna låta operatorn 
divergera, gör vi oss av med dem genom att förfina cell-indelningen ända tills 
alla noder sammanfaller med ytornas korsningspunkt i indelningen i 3-celler. På denna nivå fungerar regulariseringen precis enligt
\be
V_B = \lim_{N \rightarrow \infty} V_B^N,
\ee       
med $V_B^N = \sum_{I = 1}^N V_B^I$, där $V_B^I$ endast innehåller en nod som befinner sig i punkten för korsningen av de tre ytorna
för $V_B^I$. Tyvärr innehåller denna operator ännu en tvetydighet. Den kommer av beräkningen av $V_B$, vilken beror på
hur stigarna korsar ytorna. Nämligen om de är ovanför eller under ytan i 3-cellen i vilket fall de har en faktor $+1$ respektive $-1$ som vi inte känner till. 
Detta sköter man genom att ta ett vägt medelvärde av alla möjligheter. 

Man kan tolka spinn-nätverken geometriskt så, att de bär på kvanta av rum på varje nod av spinn-nätverket \cite{rovsm}. Dessa kvanta av rum har ingen 
definitiv form, p.g.a. att operatorerna för dem inte kommuterar, men dessa kvanta kan ''vävas'' ihop till en klassisk geometri \cite{weave}. Denna information är fullkomligt 
bakgrundsoberoende och kan användas för att se spinn-nätverken som en polymerlik kvantisering av geometrin \cite{mink}, men denna bild gäller endast 
vid det klassiska gränsvärdet av teorin och är förstås inte exakt i kvantteorin. Det viktigaste är att notera att endast de kombinatoriska aspekterna av grafen för spinn-nätverket är betydande: vilka noder är ihoplänkade med vilka noder samt areakvanttalen på stigarna och volymkvanttalen på noderna. Detta är en klar konsekvens av grundprincipen bakom SKG och spinn-skum modellerna, bakgrundsoberoendet.

Ett intressant alternativ är att tolka spinn-nätverken som sk. vridna geometrier \cite{twgeo} (eng. \emph{twisted geometries}). I detta fall tolkas en punkt i fasrummet av teorin som en ansamling diskreta geometrier m.h.a. ett antal t.ex. areor och dihedrala vinklar (eng. \emph{dihedral angle}). Dessa variabler beskriver trianglarna i den 
duala trianguleringen av spinn-nätverket m.h.a. triangelns orienterade area, två normaler till dess yta som sett från de två polyedrerna som delar på denna triangel 
och en vinkel som relateras till den extrinsiska krökningen. 

Det bör nämnas att ett godtyckligt tillstånd av geometrin inte beskrivs av ett spinn-nätverk, utan av en lineär superposition av dem. Därför måste man, för att 
dra nytta av ovanstående diskussion inför en klassisk tolkning av denna kvantgeometri, kunna betrakta koherenta tillstånd av geometrin. 
Dessa får sina största värden på klassiska intrinsiska, extrinsiska eller båda, geometrier och har konstruerats och studerats i bl.a. \cite{thm1, thm2, thm3, thm4, thiii1, qutet, twi1, eugp, biaspe}. Det är dessa, som ger oss en kontakt med den klassiska geometrin som kvanttillstånden beskriver.  

\subsection{Hamiltonbegränsningen \label{hah}}

Den sista begränsningen 
\be
H = {E^a_iE^b_j\over \sqrt{\det E}} \big({1\over 2}\epsilon^{ij}_{\, \, \, \, \, k} F^k_{ab} - (1 + \gamma^2)K^i_{[a}K^j_{b]} \big), \label{hamham}
\ee
Hamiltonbegränsningen är den svåraste att kvantisera, men som tur går det att konstruera en kvantisering av den \cite{th0, th1, th3, th2}. Denna kvantisering 
startar ifrån de följande observationerna 
\begin{align}
\epsilon^{ijk}\epsilon_{abc}{E^a_iE^b_j\over \sqrt{\det E}} &= {1\over 32\pi l_p^2\gamma}\{A_c^k, V_\Sigma \} \label{1st} \\
K^I_aE^a_I &= \{H_E, V_\Sigma \}, \label{2nd}
\end{align}
där $V_\Sigma$ är volymen av $\Sigma$ och $H_E =  {1\over 2}{E^a_iE^b_j\over \sqrt{\det E}}\epsilon^{ij}_{\, \, \, \, \, k} F^k_{ab}$ och kallas för den Euklidiska delen 
av begränsningen. Nu kan vi ändra Poison-kommutatorerna till kvant-kommutatorer bara vi klarar av att promovera Poison-kommutatorernas argument till
operatorer. Volymen har vi redan kvantiserat, men krökningen $F^k_{ab}$ har vi inte kvantiserat. Detta kan göras via en regularisering av 
holonomin för en infinitesimal slinga $\sigma$ med arean $\epsilon^2$ enligt
\be
h_{\sigma_{ab}}[A] - h^{-1}_{\sigma_{ab}}[A] = \epsilon^2F^i_{ab}\tau_i + \mathcal{O}(\epsilon^4). 
\ee
Genom en liknande procedur kan Poison kommutatorn $\{A_c^k, V_\Sigma \}$ regulariseras som
\be
h^{-1}_{e_a}[A]\{h_{e_a}[A], V_\Sigma\} = \epsilon\{A^i_a, V_\Sigma\}, \label{4th}
\ee
med $e_a$ en stig längs koordinaten $a$ av koordinatlängd $\epsilon$. Med informationen i \eqref{1st}-\eqref{4th} kan vi regularisera 
Hamiltonbegränsningen \eqref{hamham} och ange den i holonomi-flödes variabler. Denna definition av Hamiltonbegränsningen 
kommer inte att bero av regulatorn $\epsilon$, vilken kommer att försvinna i det slutliga uttrycket för den kvantiserade Hamiltonbegränsningen (se \cite{thiie} för 
en klar förklaring till hur regulatorn tas bort). Eftersom vi inte lär oss mer av att ange hela uttrycket för operatorn för Hamiltonbegränsningen, nöjer vi oss med att 
beskåda $H_E$, som i kvantiserad form är
\begin{align}
\wh{H}_E(N) =  \lim_{\epsilon \rightarrow 0}\sum_I N_I \epsilon^{abc}\Tr\Big[(\wh{h}_{\sigma^I_{ab}}[A] - \wh{h}^{-1}_{\sigma^I_{ab}}[A])\wh{h}^{-1}_{e^I_c}[A]\{\wh{h}_{e^I_c}[A], \wh{V}_\Sigma \}\Big], \label{heuclid}
\end{align}
där $N_I$ är förlopps-funktionen (eng. \emph{lapse function}) för varje cell $I$. Man kan ta bort regulatorn $\epsilon$ på följande sätt: eftersom 
$\mathcal{H}^{\mathrm{Diff}}_{kin}$ är invariant under diffeomorfismer, kommer
de nya stigarna som $\wh{H}_E(N)$ skapar att kunna placera sig var som helst, så länge de är i kontakt med noden som $\wh{H}_E(N)$ agerar på. 
P.g.a. diffeomorfisminvariansen kommer alltså placeringen av stigarna inte att ha någon betydelse. Eftersom regulatorn $\epsilon$ är längden på 
de nya stigarna anger de också platsen på spinn-nätverket för den nya stigen. I $\mathcal{H}^{\mathrm{Diff}}_{kin}$ har denna
plats ingen betydelse och regulatorn kan tas bort. M.a.o. är den inre produkten \eqref{diffprod} som projicerar ut de diffeomorfisminvarianta tillstånden 
\be
([\psi_{f, \gamma}]|\wh{H}_E(N)|\psi_{g, \gamma'}\rangle = \lim_{\epsilon \rightarrow 0}([\psi_{f, \gamma}]|\wh{H}^\epsilon_E(N)|\psi_{g, \gamma'}\rangle, 
\ee     
där $\wh{H}^\epsilon_E(N)$ representerar den regulariserade Hamiltonbegränsningen, på detta sätt väl-definierad för alla tillstånd 
$\psi_{f, \gamma} \in \mathcal{H}_{kin}$. 

Operatorn $\wh{H}_E(N)$ agerar endast på noderna av ett spinn-nätverk. Detta är i princip en konsekvens av existensen av volymoperatorn $\ha{V}_{\Sigma}$
i $\wh{H}_E(N)$. Aktionen av $\wh{H}_E(N)$ skapar nya stigar från och till en nod p.g.a. de kvantiserade holonomierna i $\wh{H}_E(N)$. Amplituden
av dessa nya stigar beror i sin tur på hur volymoperatorn agerar, vilket i sin tur beror på lokala detaljer av noden som t.ex. hur många stigar som 
lämnar och kommer till noden och deras spinn. De nya stigarna som $\wh{H}_E(N)$ adderar till noden kommer att skapa 
nya 3-valenta noder (noder med tre stigar) som en ny aktion av Hamiltonbegränsningen kommer att annihilera. D.v.s. kommutatorn av två Hamiltonbegränsningar 
kommer att vara 0 enligt
\be
([\psi_{f, \gamma}]|[\wh{H}_E(N), \wh{H}_E(M)]|\psi_{g, \gamma'}\rangle = 0, \label{cloH}
\ee       
men endast då vi betraktar diffeomorfisminvarianta tillstånd \cite{th3}. D.v.s. begränsnings\-algebran kommer inte generellt sett att sluta sig enligt
\be
[\wh{H}_E(N), \wh{H}_E(M)] = \wh{V}(N, M), \label{stclo}
\ee
där $\wh{V}(N,M)$ är en operator som är proportionell mot diffeomorfismgeneratorn. 

Det verkar klart att \eqref{stclo} inte kan nås i den nuvarande 
formuleringen av SKG, eftersom det inte finns någon infinitesimal operator för diffeomorfismer i SKG (se sektion \ref{diffeo}). Detta är ett av problemen 
med kvantiseringen av Hamiltonbegränsningen. Eftersom $[\wh{H}_E(N), \wh{H}_E(M)]$, de andra kommutatorerna i begränsningsalgebran är problemfria, 
endast sluter sig under kravet \eqref{cloH}, vilket vi håller fast vid i SKG, kommer vi att
introducera en stor grad av tvetydighet för tillstånden i $\mathcal{H}_{phys}$. Dessa är bl.a. att man kan välja holonomierna i \eqref{heuclid} i en annan representation 
än den fundamentala för SU(2). Detta har en effekt i SlingKvantKosmologi \cite{Bojo}, stigarna i holonomierna för \eqref{heuclid} kan väljas på många olika sätt vilket också 
inför tvetydigheter i teorin \cite{Ashrev} och dessutom kommer vi att måsta fundera över ordningen på operatorerna i \eqref{heuclid} som inför en annan grad
av tvetydighet i SKG. Vi har alltså väldigt många tvetydigheter i SKG vilka alla mer eller mindre rör sig kring kvantiseringen av Hamiltonbegränsningen, 
och står i stark relation till att begränsningsalgebran endast sluter sig enligt \eqref{cloH}. Det bör dock nämnas att det går att skapa lineära kombinationer 
av spinn-nätverk som lösningar på Hamiltonbegränsningens aktion på diffeomorfisminvarianta spinn-nätverk \cite{thiie}, men dessa lösningar är endast 
formella eftersom koefficienterna för de lineära kombinationerna tillsvidare är okända. 

\subsection{SKG:s status just nu \label{sttnw}}

Vi har här sett att de första och nästan hela det andra steget i Dirac-programmet kunnat utföras inom SKG, trots att den Hamiltonska begränsningen ställer till med problem. 
Det tredje steget, att hitta lösningsrummet för Hamiltonbegränsningen är problematiskt, p.g.a. dess väldigt tvetydiga natur. Det fjärde och sista steget i Dirac programmet, 
att hitta en fullständig mängd gaugeinvarianta Dirac observabler, har inte heller kunnat utföras utan en bra förståelse av den Hamiltonska begränsningen, men man 
har lärt sig en del om dessa observabler trots det (för en review se \cite{tamb}), t.ex. konstruktionen av en fullständig mängd observabler för gaugeinvarianta 
bakgrundsfria och därmed icke-störningsteoretiska leksaksmodeller \cite{obse1, gies, obse2, domag}, samt en klar tanke om hur de borde konstrueras inom allmän 
relativitet \cite{rov, rove2}.  

Förutom denna konstruktion av SKG, har andra resultat också uppnåtts inom strukturen för teorin. T.ex. en tanke om hur tiden återuppstår i teorin. Denna 
tid kallas termodynamisk tid \cite{thmtm, thmtm1} och som namnet säger är tid i denna tolkning en egenskap som uppstår vid ett termodynamiskt gränsvärde av den totala
teorin. M.a.o. blir tiden en sorts makroskopisk observabel av statistisk natur. Det bör nämnas att den fulla teorin inte behöver denna konstruktion eftersom
teorin förutsäger korrelationer mellan fysikaliska variabler, d.v.s. hur de förhåller sig till varandra och inte i förhållande till en tid så som vi vanligtvis tänker på 
fysikaliska förutsägelser, men oberoende, är tanken intressant. Dessutom har betraktelsen av denna termodynamiska tid, också lett till en intressant utveckling gällande
en möjlig konstruktion av en statistisk mekanik som innehåller gravitation där den termodynamiska tiden relaterar frihetsgraderna för modellen till Shannon informationens konservering \cite{thmtm2}.

En annan intressant utveckling är Kodama tillståndet, som tänktes vara vakuumtillståndet för SKG med en 
positiv kosmologisk konstant \cite{kod, smo}, men detta har ifrågasatts av \cite{witt} med argumentet att tillståndet skulle leda till negativa energier. I vilket fall som helst är 
Kodama tillståndet en exakt lösning på alla begränsningar i SKG i självduala variabler ($\gamma = \pm i$) och därför mycket intressant med tanke på de gaugeinvarianta observablerna.    

Tvetydigheterna i Hamiltonbegränsningen är definitivt ett problem, men vi måste påminna oss om att materia inte spelar någon roll i aktionen 
\eqref{holst} för SKG, vilket den säkert på fundamental nivå gör\footnote[1]{Det går att konstruera SKG med den kvantfältteoretiska standardmodellen som materia \cite{th2}, men 
detta är inte den fundamentala nivå som menas här. P.g.a. Einstein ekvationernas koppling mellan geometri och materia,
är det lätt att tänka sig att de förenas på en djupare kvantgravitationell nivå och indelningen i två olika strukturer, materia och geometri är endast en 
produkt av vår ovetskap om vad kvantgravitation är. Se också \cite{boj}.}. En annan svaghet av SKG är att vi kvantiserar i ett partiellt gauge-mått av Lorentzgruppen (SU(2)). Det 
skulle vara bättre att hålla fast vid hela Lorentzgruppen som en lokal symmetri för kvantgravitation, för att bättre kunna upptäcka teorins struktur. 
Detta är bl.a. en av tankarna bakom spinnskum modellerna, men den främsta är att lösa Hamiltonbegränsningen. Man kan, formellt, skapa en 
expansion av projektioner på Hilbertrummet $\mathcal{H}_{phys}$ som skapar en inre produkt på $\mathcal{H}_{phys}$ m.h.a. abstrakta 
spinn-nätverk. Denna expansion har en klar analogi till Feynmans stig-integraler och vi kommer att se närmare på den i sektion \ref{sskum}.  

\subsection{Svarta hål i SKG \label{SHSKG}}

P.g.a. termodynamiken för svarta hål (se \cite{wald} för en review) och singulariteten innanför händelsehorisonten, är svarta hål mycket intressanta objekt att
studera inom alla teorier för kvantgravitation. Minimikraven på en kvantgravitationsteori är att singulariteten borde försvinna och 
speciellt att frihetsgraderna bakom Bekestein-Hawking lagen om area, via entropin för det svarta hålet, kan identifieras.  

Inom SKG kan man studera svarta hål genom att starta från Holst aktionen \eqref{holst} på ett mångfald med en yttre gräns (eng. \emph{boundary}).
Denna gräns kommer till p.g.a. händelsehorisonten för det svarta hålet och kallas för en isolerad horisont (se \cite{iso} för en review). En isolerad horisont är 
definierad med tanke på isolerade jämviktstillstånd i termodynamik och detta resulterar i att det svarta hålet inte växelverkar med sin omgivning, trots att omgivningen i sig 
självt kan vara dynamisk. Man kan tänka på dem som en generalisering av Killing horisonter. Medan Killing vektorfältet är definierat i en lokal omgivning av horisonten,
definieras den isolerade horisonten endast m.h.a. den intrinsiska och extrinsiska geometrin av horisonten. De isolerade horisonterna satisfierar 
en lokal termodynamisk första lag \cite{alex3} och därför kan man mycket väl studera entropin för svarta hål med isolerade horisonter.

Med den föregående definitionen på isolerad horisont kan vi snabbt stöka oss igenom hur de svarta hålen behandlas inom SKG. Historiskt 
gavs de första idéerna till denna konstruktion i \cite{rokra}.
Vi börjar med att definiera den yttre gränsen av mångfaldet $\Delta$ topologiskt som $\Delta = \mathbb{R} \times S^2$. Sedan gör vi 3+1 ADM analysen 
med Ashtekar variablerna, precis som i sektion \ref{hamam}, men nu har mångfaldet $\Sigma$ en yttre gräns. Efter det kräver vi att
den yttre gränsen $\Delta$ är en isolerad horisont, vilket ger oss gränsvillkor på fälten $E$ och $A$ som definierades i den Hamiltonska 
analysen med Ashtekar variabler. Den symplektiska strukturen som vi åstadkommer i denna modell kvantiseras så, att
fälten innanför horisonten kvantiseras separat från fälten ovanpå horisonten. 

Fälten som kvantiseras innanför horisonten 
kvantiseras likadant som i sektion \ref{skkg}, med en ny aspekt vilken är att stigarna kan sluta på horisonten. Dessa stigar karakteriseras 
av spinn kvanttal $m_p$ och $j_p$, där $j_p$ beror av representationen av spinnet på stigen.  

Fälten som kvantiseras ovanpå horisonten 
har en yt-term som är termen för en SU(2) Chern-Simons teori \cite{bhcs}. Konnektionen för denna Chern-Simons teori är lokalt platt (eng. \emph{flat}) 
men den har frihetsgrader där stigarna punkterar horisonten. Dessa beskrivs, grovt sagt, av kvanttal $m_p'$ och $j_p'$, vilka är liknande som kvanttalen 
som beror av spinnet på stigen. 

Då man kvantiserar gränsvillkoren för fälten $E$ och $A$ får man en operatorekvation, vars lösningar 
är tensorprodukter av tillstånd innanför och ovanpå horisonten, där kvanttalen $m_p = m_p'$ och $j_p = j_p'$. Om man sedan fixerar 
den makroskopiska arean på det svarta hålet att vara $a_0$, kommer detta att slå fast antalet punkturer i horisonten och de kan räknas, vilket ger oss
entropin
\be
S(a_0) = \ln(N(a_0)) = {\gamma\over \gamma_0} {a_0\over 4\pi l_p^2} + \mc{O}(\ln {a\over l_p^2}), \label{bhS}
\ee    
där $\gamma_0 = 0.2375...$ konstanten följer från beräkningen av antalet punkturer \cite{dom, mire, mupp}. Denna formalism tillåter också beräkningen av  en logaritmisk 
korrektion \cite{mopp}. M.a.o., om vi fastställer Barbero-Immirzi parametern som 
$\gamma = \gamma_0$, har vi hittat Bekestein-Hawking lagen om arean för ett svart hål. Denna beräkning kan göras för vilket som helst 
svart hål i Kerr-Newman familjen och resultatet är det samma \cite{kerr, new}. Beräkningen är oberoende av detaljerna gällande Hamiltonbegränsningen. 

Trots att resultatet \eqref{bhS} endast verkar uppnås för $\gamma = \gamma_0$, är detta inte det enda värdet på $\gamma$ som ger oss
Bekestein-Hawking formeln. Om vi tillåter att $\gamma$-parametern
också antar komplexa värden, kan vi analytiskt fortsätta (eng. \emph{analytically continue}) beräkningen på ett sådant sätt att resultatet \eqref{bhS} 
också uppnås ifall $\gamma = \pm i$ \cite{alex2}, vilket återförenar oss med de gamla självduala Ashtekar variablerna.    

\section{SlingKvantKosmologi (SKK)}

Kosmologi ser ut att vara den enda arenan där en teori för kvantgravitation kan testas experimentellt. Detta, eftersom acceleratorer inte kommer ens i närheten 
av de energier som krävs för Plancklängd och de starkaste anhopningarna massa vi känner till sker i svarta hål och i universums början, Big Bang. 
Därför är det viktigt för varje teori av kvantgravitation att kunna passa ihop med de experiment som ger åt oss den nutida bilden av en Big Bang kosmologi. 
Inom SKG förväntar man sig att den kosmologiska sektorn av teorin skall inbegripa SlingKvantKosmologi (SKK)\footnote[1]{Denna reviewartikel är inte en review av SKK 
(se t.ex. \cite{Ashie, mer} för detta). SKK nämns endast p.g.a. dess viktiga position inom SKG och för att ge en idé om vilka delar av SKG som används inom SKK.} i en 
eller annan form. SKK består av symmetrireducerade modeller, i vilka symmetrin reduceras före kvantisering, som använder sig starkt av metoder, tekniker och resultat från SKG. Detta leder till de stora resultaten inom SKK vilka bl.a. är en resolution av singulariteten i Big Bang i en Big Bounce samt tillräcklig inflation för att skapa vårt nutida universum i nästan alla modeller av SKK. 

I det följande skall vi se lite närmare på strukturen av SKK. Vi börjar med att introducera den homogena FLRW ($k=0, \Lambda = 0$) modellen i delsektion \ref{flrwk}. Sedan 
noterar vi hur man bygger effektiva modeller inom SKK i delsektion \ref{eff} och i delsektion \ref{apprx}, ger vi en motivering till varför symmetrireduceringen, trots att den 
görs på den klassiska rymden, kan ge oss betydande resultat.  

\subsection{Homogen FLRW kosmologi \label{flrwk}}

Id\'en bakom SKK baserar sig på minisuperrymder \cite{wheel} (eng. \emph{minisuperspaces}), symmetrireducerade modeller av gravitation. Vi presenterar här i korthet
strukturen av kvantiseringen av FLRW ($k=0, \Lambda = 0$) modellen inom SKK, eftersom den är grundstenen i byggandet av alla minisuperrymds modeller inom SKK. Kvantiseringen av den homogena FLRW modellen påbörjades i \cite{bojo1, bojo2}, speciellt resultatet att denna modell inte innehöll
någon singularitet \cite{bojo1}, ledde till mycket aktivitet inom området. Efter dessa resultat byggdes modellens kinematik upp på ett mera stabilt plan \cite{ashkin} 
efter ny utveckling inom SKG och detta ledde till slut till kvantiseringen av denna modell med ett masslöst skalärt fält som materiekälla samt studierna 
av dess kvantdynamik \cite{mass, mass2, mass3, mass4}. 

I all enkelhet startar man från att begränsa sig till den Euklidiska gruppen 
$\mc{S}$. I detta fall agerar den 3-dimensionella gruppen för translationer $\mc{T}$ (en undergrupp till $\mc{S}$), som försäkrar oss om att vi har en homogen modell, 
transitivt på mångfaldet 
$\Sigma$. Detta betyder i all korthet att topologin på mångfaldet $\Sigma = \mathbb{R}^3$. Det essentiella är att Lie-algebran för translationsdelen av 
den Euklidiska gruppen kommer att ha en ekvivalensklass av positivt definita metriker, som alla är relaterade via en konstant. Dessa är våra FLRW-metriker.  

Eftersom beräkningen av integraler i den icke-kompakta $\mathbb{R}^3$ leder till oändligheter, introducerar man en cell $\mc{V}$ som vi anpassar 
efter triaderna för en metrik $\dot{q}_{ab}$ som vi fixerar. De fysikaliska slutsatserna är givetvis helt oberoende av valet av denna cell. Cellen kan 
väljas kubisk i relation till den fixerade metriken och diagonal 
$\dot{e}^i_a = \delta^i_a$\footnote[1]{P.g.a. att modellen vi betraktar är homogen är det tillräckligt att betrakta en cell, eftersom vad som händer i hela universum 
kan extrapoleras ifrån denna.}. Introduktionen av cellen $\mc{V}$ är i enlighet med SKG där vi regulariserar t.ex. Hamiltonbegränsningen och area- och 
volymoperatorn enligt samma idéer. Den fixerade metriken kommer från
\be
\dot{q}_{ab} dx^adx^b = dx^2_1 + dx^2_2 + dx^2_3, \, \, \,  \mathrm{d"ar}\, \, x^a \in [0, l_o]\, \mathrm{med}\,  l_o\,  \mr{fixerad}, 
\ee  
där $\dot{q}_{ab}$ är den fixerade 3-metriken och $V_o = l_o^3$ är volymen för cellen $\mc{V}$. Den fysikaliska metriken är relaterad till den 
fixerade metriken m.h.a. faktorn för skala $a$ (eng. \emph{scalefactor}) enligt $q_{ab} = a^2\dot{q}_{ab}$. M.a.o. fixerar detta diffeomorfism invariansen 
i hela SKG till denna cell. Utöver detta väljer vi variabler $(\ti{A}_a^i, \ti{E}^b_j)$ så att vi för alla $s \in \mc{S}$ har variabler som är invarianta under 
Gauss-begränsningen. Detta val och ett val av bättre variabler för kvantisering \cite{ashkin} leder oss till den symplektiska strukturen
\be
\{c, p\} = {8\pi G \gamma \over 3}, \, \, \, \mr{med}\, \ti{A}_a^i = cV_o^{-1/3} \dot{e}^i_a, \, \mr{och}\, \ti{E}^b_j = pV_o^{-2/3}\sqrt{\dot{q}}\dot{e}_j^b,  
\ee  
där de nya kanoniska variablerna är $c$ och $p$, variabler med en punkt ovanför är fixerade m.h.a. den fixerade metriken 
och $\sqrt{\dot{q}}$ betecknar dess determinant. Med dessa val av gauge och variabler har vi begränsat en teori med ett oändligt antal 
frihetsgrader till ett ändligt antal, orsaken vi benämner minisuperymdsmodellerna inom SKK, symmetrireducerade modeller. 

Nu kan vi slutligen införa ett skalärt masslöst fält $\phi$ som materian i denna FLRW modell. Denna har den symplektiska strukturen 
$\{\phi, P_{\phi}\} = 1$, där $P_\phi$ är dess kanoniska rörelsemängd. Detta ger oss den totala Hamiltonbegränsningen $C_H$ för denna modell som 
\be
C_H = C_{grav} + C_{mat} = -{6\over \gamma^2}c^2\sqrt{|p|} + 8\pi G {P_\phi^2 \over |p|^{3/2}} = 0, \label{cham}
\ee 
där $ |p|^{3/2} = V$ är den fysikaliska volymen för cellen $\mc{V}$. Den fysikaliska volymen $V$ beror inte av fixeringen av metriken.  

Tanken i SKK är att kvantisera enligt id\'eerna inom SKG och då behöver vi övergå till holonomi-flödes variablerna. Eftersom modellen 
är homogen kan vi fritt orientera triaderna längs den fixerade cellens sidor med den orienterade längden $\mu V_o^{1/3}$, där 
$\mu$ är ett reellt tal. Holonomin längs en sida $i$ blir då 
\be
h^{\mu}_i(c) = e^{\mu c \tau_i} = \cos({\mu c\over 2})\id + 2\sin({\mu c\over 2})\tau_i
\ee
och flödet fortsätter att beskrivas av variabeln $p$. Från detta drar vi slutsatsen att konfigurationsalgebran för den gravitationella 
delen av modellen genereras av $\mc{N}_\mu(c) = e^{{i\over 2}\mu c}$. Denna algebra är algebran för cylindriska funktioner 
Cyl$_\mc{S}$, i analogi med algebran för cylindriska funktioner i SKG. 

Som nästa steg \`a la SKG, väljer vi representation för denna algebra där $\mc{N}^\mu(c)$ och $p$ representeras av operatorer så 
att operatorn $\ha{p}$ inte är kontinuerlig i konnektionen och således finns det ingen operator som representerar $c$ \cite{ashkin}. Detta är analogt med 
SKG där vi inte har någon operator för konfigurationsvariabeln $A^i_a$. På detta sätt blir kvantkonfigurationsrummet Bohr-kompaktifieringen 
av den reella linjen, $\mathbb{R}_{Bohr}$ och Haar måttet för denna rymd är Bohr måttet \cite{bohrme}. Eftersom denna rymd är isomorf till en rymd där
funktioner av $\mu \in \mathbb{R}$ är kvadratiskt summerbara i Bohr måttet kommer vi att kunna ange en bas av tillstånd $ |\mu\rangle$ för modellen 
med egenskapen
\be
\ha{\mc{N}}_{\mu'}(c)|\mu\rangle = |\mu + \mu'\rangle,  
\ee  
där det kinematiska rummet är Cauchykompletteringen av Cyl$_\mc{S}$ med tanke på den inre produkten $\langle \mu|\mu '\rangle = \delta_{\mu\mu'}$. 
Den andra kanoniska variabeln i detta Hilbertrum, som hädanefter kallas $\mc{H}_{grav}$, agerar på dessa tillstånd enligt
\be
\ha{p}|\mu\rangle = {4\pi l^2_p\gamma\over 3} \mu |\mu\rangle. 
\ee     
Hilbertrummet för materiedelen av begränsningen \eqref{cham} kan representeras enligt en standard Schrödingerrepresentation
med $\ha{P}_\phi = -i\partial_\phi$ och $\ha{\phi} = \phi$ i rummet $L^2(d\phi, \mathbb{R})$, så att det totala kinematiska Hilbertrummet är
$\mc{H}^{tot}_{kin} = \mc{H}_{grav}\otimes L^2(d\phi, \mathbb{R})$.

Efter detta måste vi ännu kvantisera Hamiltonbegränsningen \eqref{cham}. Detta gör vi precis som i SKG där vi skriver om $H$ (se sektion \ref{hah})
med Poisonkommutatorer och kvantiserar volym- och holonomioperatorerna. Faktum är att vi endast behöver fundera över kvantiseringen av krökningen 
$F^k_{ab}$, eftersom spinn-konnektionen är 0 i denna platta FLRW modell, vilket ger oss från \eqref{hamham}
\be
H_{FLRW} = -{1\over \gamma^2}\int_\Sigma d^3x{E^a_iE^b_j\over \sqrt{\det E}} {1\over 2}\epsilon^{ij}_{\, \, \, \, \, k} F^k_{ab}.
\ee  
Krökningen kan vi nu beräkna då vi väljer en kvadrat som Wilsonslinga 
\be
h^\mu_{{\Box}_{ij}} = h^\mu_i h^\mu_j (h^\mu_i)^{-1} (h^\mu_j)^{-1}
\ee
med arean $A_\Box = \mu^2V_o^{2/3}$. Enligt metoden presenterad i sektion \ref{hah} får vi då
\be
F^i_{ab} = -2\lim_{A_{\Box} \rightarrow 0} \Tr\Big({h^\mu_{{\Box}_{jk}} - \delta_{jk}\over A_\Box}\tau^i\Big) \dot{e}^j_a\dot{e}^k_b.
\ee
Detta gränsvärde är väl definierat i den klassiska teorin, men i kvantteorin divergerar det. \emph{MEN}, eftersom vi känner till areaspektret 
i SKG \eqref{kvar}, vet vi att detta gränsvärde endast kan tas till minimiegenvärdet för area inom SKG som vi kallar $\Delta$. Detta värde väljs som 
$A_{\Box_{min}} = \bar{\mu}^2V_o^{2/3}$, så att den fysikaliska arean som minst har värdet $\Delta$. Valet ger oss också minimiflödet 
$E_{\Box_{min}} = \bar{\mu}^2 p$\footnote[1]{Detta val av minimiarea brukar kallas inom SKK för förbättrad dynamik \cite{mass3} (eng. \emph{improved dynamics}). Detta beror
på att då man förut antog att $\mu$ kunde väljas som en konstant, resulterade detta i en dynamik som inte överensstämde med allmän relativitet \cite{mass2}}, 
så att vi kan definiera den regulariserade krökningen och m.h.a. den, definiera krökningsoperatorn $\ha{F}^i_{ab}$. Till slut 
gör vi ännu en kanonisk transformation till en ny bas, via variablerna
\begin{align}
b &= {\bar{\mu}c\over 2} \\
v &= {1\over 2\pi \gamma l_p^2\sqrt{\Delta}}\sgn (p)|p|^{3/2}, \label{vvv}
\end{align}
med $\{b, v\} = 1$ och $\mc{N}_{\bar{\mu}}| v \rangle = |v +1\rangle$, vilket slutligen ger oss den kvantiserade Hamiltonbegränsningen. För att slutföra kvantiseringen av 
Hamiltonbegränsningen måste man ännu välja en operatorordning i Hamiltonbegränsningens operator, frikoppla (eng. \emph{decouple}) noll volymtillståndet $|0\rangle$ samt tillstånd med olika orientering i förhållande till triaderna. Orsakerna är att Hamiltonbegränsningens operator inte är själv-adjungerad ifall den inte har en symmetrisk 
operatorordning och inte heller ifall nolltillståndet finns med i volymspektret. Att frikoppla tillstånd med olika orientering i förhållande till triaderna är inte nödvändigt, andra möjligheter har också studerats \cite{mass3, mass4, mer2, mer3, amb1, amb2} (vi väljer att följa \cite{mer2, mer3}), men det simplifierar det kinematiska Hilbertrummet genom att dela in det i superselektionssektorer som man kan studera var för sig. Själva Hamiltonbegränsningsoperatorn kommer att se ut som
\be
\ha{C}_H = -{6\over \gamma^2}\ha{\Omega}^2 + 8\pi G\ha{P}_{\phi}^2, \label{lutop}
\ee
med $\ha{P}_{\phi}^2 = -\partial_\phi^2$ och $\ha{\Omega}^2 = {1\over 4i\sqrt{\Delta}}|\ha{p}|^{3/4}\Big[(\ha{\mc{N}}_{2\bar{\mu}} - \ha{\mc{N}}_{-2\bar{\mu}}) \sgn(\ha{p})  
+ \sgn(\ha{p})(\ha{\mc{N}}_{2\bar{\mu}} - \ha{\mc{N}}_{-2\bar{\mu}})\Big] |\ha{p}|^{3/4}$. Dessa variabler är enklare att arbeta med 
eftersom $\ha{\Omega}^2$ och $\ha{P}^2_{\phi}$ blir Diracobservabler som kommuterar med $\ha{C}_H$ i denna bas. Dessutom är $v$ i \eqref{vvv} proportionell 
mot den fysikaliska volymen för cellen $|p|^{3/2}$ och detta resulterar i att $\ha{v}$:s egenvärden är proportionella mot volymens egenvärden.

Efter allt detta är vi i position att använda grupp medelvärdes (eng. \emph{group averaging}) metoder. Det fysikaliska Hilbertrummet för denna 
FLRW modell är de tillstånd som hålls invarianta under aktionen av gruppen som fås genom en själv-adjungerad utbyggnad (eng. \emph{self-adjoint extension}) 
av operatorn \eqref{lutop}. Vi hittar dem genom att ta gruppmedelvärdet över denna grupp. Ytterligare kommer denna procedur att ge oss en 
inreprodukt som ger Hilbertstrukturen åt det fysikaliska Hilbertrummet. Denna metod ger oss tillstånden
\begin{align}
\Psi(v, \phi) &= \int_0^{\infty}d\lambda e^\epsilon_\lambda(v)[\ti{\psi}_+(\lambda)e^{i\nu(\lambda)\phi} + \ti{\psi}_-(\lambda)e^{-i\nu(\lambda)\phi}] \label{stat} \\
\nu(\lambda) &= \sqrt{3\lambda\over 4\pi l_p^2\gamma^2}, 
\end{align}    
där $e^\epsilon_\lambda(v)$ är en egenfunktion av operatorn $\ha{\Omega}^2$, $\epsilon \in (0, 4]$ är relaterad till $\ha{\Omega}^2$:s egenvärden och $\pm$-beteckningen anger de olika kontributionerna i förhållande till orientationen av triaden. Den inre produkten blir
\be
\langle \Psi_1|\Psi_2 \rangle_{fys} = \int_0^\infty d\lambda[\ti{\psi}^*_{1+}(\lambda)\ti{\psi}_{2+}(\lambda) + \ti{\psi}^*_{1-}(\lambda)\ti{\psi}_{2-}(\lambda) ].
\ee
Det sista steget blir att välja Dirac observabler i detta system. Det torde vara ganska klart att vi kan och det lönar sig att välja $\phi$ som ''klockan'' för detta system
eftersom vi kan beskriva evolutionen av systemet i förhållande till $\phi$. I så fall kommer $\nu$ att spela rollen som frekvens i förhållande till denna variabel (se \eqref{stat}) och 
vi kan definiera en fullständig mängd Dirac observabler i ''evolution'' i denna model. Dessa kan väljas som $\ha{v}(\phi)|_{\phi_0}$, volymen vid tidpunkten $\phi_0$ och rörelsekonstanten$\ha{P}_{\phi}$. Detta var vårt mål: vi har utfört Dirac programmet ända till slut i denna enkla FLRW-model. 

Modellen vi skapat är fri från en Big Bang singularitet \cite{mass3}. Det finns många sätt att se detta. Vi kunde t.ex. noterat att då vi frikopplar volymtillståndet $|0\rangle$ från vår 
fysikaliska Hilbertrymd, kommer vi inte att kunna nå noll volym. Detta är dock ett lite dumt konstaterande eftersom det ger känslan att vi gjort oss av med singulariteten 
för hand. Ett bättre sätt att se att själva $|0\rangle$ tillståndet inte behövs är att notera att egenfunktionerna $e^\epsilon_\lambda(v)$ i \eqref{stat} delar sig i två
grenar då $v$ är stor. En gren egenfunktioner som kontraherar och en som expanderar och de kan tolkas som ingående och utgående vågor. Eftersom båda 
deltar med en lika stor amplitud måste den fysikaliska lösningen \eqref{stat} beskriva ett universum som kontraherar och via en Big Bounce börjar expandera igen utan att gå via
noll volymtillståndet, eller vice versa. På detta sätt undviks Big Bang singulariteten i denna, enklaste FLRW modell.

Denna modell kan också utvidgas för att studera ett slutet FLRW universum och ett öppet FLRW universum (se t.ex. \cite{Ashie} för detaljerna och referenserna 
gällande alla dessa modeller). I det slutna fallet ($k = +1$) får vi istället 
för en Bounce händelse, en cyklisk modell där universat expanderar och kontraherar periodiskt. I modellen för det öppna universat ($k = -1$) har vi endast 
en Bounce händelse, om vi startar från ett kontraherande universum, och volymen blir oändlig för det expanderande universat. Precis som väntat. Modellen 
kan också studeras både för en positiv och en negativ kosmologisk konstant. I det positiva fallet, har Hamiltonbegränsningsoperatorn många olika 
själv-adjungerade utbyggnader, men efter ett visst gränsvärde $\Lambda > \Lambda*$ som är av ordningen Planckenergi, finns det, trots att Hamiltonbegränsningen 
fortfarande är själv-adjungerad, inga intressanta fysikaliska tillstånd i det fysikaliska Hilbertrummet\footnote[1]{Maximalt, innehåller detta Hilbertrum lösningar 
som är icke-triviala men som sönderfaller alldeles för snabbt för att beskriva vårt universum.}. Då den kosmologiska konstanten å andra sidan är 
negativ, kommer också fallet $k = -1$ att vara cykliskt medan modellerna $k = 0, +1$ fortsätter att vara det, liksom i $\Lambda = 0$ modellen.   

\subsection{Effektiv SKK \label{eff}}

Förutom denna typs dynamik där vi söker den inre produkten för det fysikaliska Hilbertrummet, kan man också konstruera effektiva ekvationer 
för SKK \cite{effskk, effskk2, va}. Dessa kan konstrueras då man har materia i modellen som man kan använda som intern klocka. D.v.s. man beräknar den Hamiltonska 
evolutionen på klassisk nivå i förhållande till det skalära fältet, innan kvantisering och sedan använder man metoder ifrån geometrisk kvantmekanik
för att approximera den Hamiltonska evolutionen för $1 \gg v$, m.a.o. stora volymegenvärden. Denna metodik ger oss en sorts effektiv Hamiltonsk begränsning 
\cite{va, boj1, boj2} vilken i sin tur leder till Friedmannekvationen och Hubbleparametern som
\begin{align}        
{1\over 9}({\dot{v}\over v})^2 &\equiv H^2 = {8\pi \over 3}\rho(1 - {\rho\over \rho_{crit}}) \label{fried} \\
H &=  {1\over 2\gamma \lambda} \sin(2\lambda b),
\end{align}
där $\lambda^2 = \Delta$, $\rho_{crit} = {3\over 8\pi\lambda^2\gamma^2}$ och $v$ och $b$ definierades i föregående sektion. Vi ser att då 
$1 \gg \lambda b$ eller $\rho_{crit} \gg \rho$ får vi tillbaka den bekanta ''klassiska'' Friedmannekvationen. Modifikationerna till 
Friedmannekvationen och Hubbleparametern kommer direkt ifrån kvantgeometrin i SKK trots att vi givit den i formen \eqref{fried}. Detta är klarare då 
vi skriver om ekvationen och väljer som materia ett skalärt fält med en potential $V(\phi)$, vilket ger oss
\be
{\sin^2\lambda b\over \lambda^2\gamma^2} = {8\pi\over 3}(\frac{1}{2}\dot{\phi}^2 + V(\phi)).
\ee
M.h.a. av denna typs effektiva ekvationer kan man studera inflation inom SKK, då vi befinner oss inom approximationen $1 \gg v$. Det har visat sig 
att SKK producerar inflation också utan ett skalärt fält genom kvantgeometriska effekter. Men denna sk. superinflation producerar inte tillräckligt 
med $e$-veck (eng. $e$-\emph{foldings}) för inflation och därför brukar man med dagens 
kunskap om SKK säga att det behövs ett massivt skalärt fält för att skapa den inflation som behövs för vårt nutida universum att existera. Om man inkluderar 
detta fält i SKK är resultatet att SKK oberoende av intialvillkor och kommer att skapa tillräckligt med inflation för vårt nutida universum \cite{prob}. Mera exakt 
sagt är resultatet, då man bestämmer början på inflationen som den tidpunkt då universat börjar sin Big Bounce, vi går från kontraherande till expanderande 
universum, i en modell med ett skalärt fält i en kvadratisk potential och väljer att se på alla fältkonfigurationer i fasrummet som befinner sig utanför den 
7 åriga WMAP datans fasrum, att universat kommer att nå fasrummet för WMAP datan med en sannolikhet på $1 - 3\times10^{-6}$. D.v.s. tillräcklig inflation 
uppnås nästan säkert inom SKK.  
  
\subsection{Hur pålitlig är SKK? \label{apprx}}

De fina resultaten inom SKK, upplösningen av singulariteterna, de cykliska modellerna och tillräcklig inflation är alla producerade inom en modell
som är symmetrireducerad på en klassisk nivå som vi sedan kvantiserar enligt SKG. Man borde givetvis i detta fall ställa sig frågan, hur tillförlitliga 
är dessa resultat, då vi inte hittat dem från en fullständig teori med ett oändligt antal frihetsgrader, utan endast använt ett ändligt antal av dem? 
Det står klart att de borde tas med en sund mängd skepsis. Utan kännedom av det fysikaliska Hilbertrummet i SKG är denna fråga omöjlig att svara
på med säkerhet, men det finns exempel inom fysik som stöder tanken att SKK mycket väl kan vara den kosmologiska sektorn av SKG. Följande tre argument 
kan hittas i \cite{Ashie}.  

Vi börjar med ett hypotetiskt exempel. Tänk om vi hade kvantelektrodynamik i sin fulla glans utan att känna till en beskrivning av väte atomen. Då kunde 
någon föreslå som en simplifierad modell att starta från sfärisk symmetri, behandla protonen och elektronen som en partiklar och sedan kvantisera detta 
system. Modellen skulle m.a.o. frysa ut alla strålningsmoder (eng. \emph{radiative modes}) så att vi har en form av kvantmekanik: Diracmodellen
av väteatomen. Symmetrireduktionen känns så drastisk att man väldigt fort skulle reagera med att säga att det är mycket osannolikt att denna modell beskriver
väteatomen tillräckligt väl. Trots det, vet vi att den gör det.  

Som ett annat exempel kan vi ta singulariteterna i allmän relativitet. I början tänkte man sig att dessa endast var en konsekvens av den höga symmetrin 
i ekvationerna och att mera realistiska former av materia skulle göra sig av med dem. Detta visade sig inte vara fallet i.o.m. singularitetsteoremen av Penrose, 
Hawking, Geroch och andra. Alltså kan vi konstatera att konsekvenserna av de symmetrireducerade modellerna var korrekta gällande singulariteterna.    

Slutligen finns det faktiskt en betraktelse över alla andra. I \cite{Kuch} byggde man in en minisuperrymd A i en annan som vi kallar B och kvantiserade de
olika teorierna och fann att A, som klassiskt motsvarade en sektor i B, inte längre gjorde det efter kvantisering av båda. Problemet i denna analys 
var att de extra frihetsgraderna i B frystes bort, de integrerades inte bort. Om man integrerar bort dessa frihetsgrader, har det visats att om A är en k=0 FLRW modell och 
B en Bianchi I modell med samma materia, kommer den klassiska sektorn i B att motsvara A både på kvant och klassisk nivå \cite{ased}, så som man vill att
SKK skall representera den kosmologiska delen av SKG. D.v.s. sensmoralen är att de symmetrireducerade modellerna kan mycket väl beskriva 
delar av SKG, så länge man gör reduceringen på ''rätt sätt''. Diskussionen om vilket är ''rätt sätt'' är givetvis inte alls enkel och inget definitivt svar kan ges 
förrän hela lösningsmängden för $\mc{H}_{phys}$ inom SKG är känd. 

\section{Spinnskum Modeller \label{sskum}}

Inom SKG var dynamiken det stora frågetecknet för teorin. Hamiltonbegränsningen var tvetydig och svår att tolka fysikaliskt med dess egenskap att skapa 
tre-valenta noder utan volym. Eftersom detta problem verkar svårt att komma åt inom strukturen för SKG, kunde man tänka sig att konstruera en kvantgravitationsteori 
i vilken man behåller de bästa delarna av SKG (spinn-nätverken), medan man ger upp andra. Detta är tanken bakom spinnskum modellerna, vilka är ett försök till en 
kovariant regulariserad stig-integral formulering av en teori för kvantgravitation. De är en sorts fortsättning på det arbetet och de idéer som introducerades av Hawking 
\cite{hawk} och Hartle och Hawking \cite{hahaw}. 

För att introducera dessa modeller, tar vi först en titt på Regge gravitation i delsektion \ref{rggra}, vilket är en 
diskretiserad formulering av allmän relativitetsteori som är mycket central för konstruktionen av spinn-skum modellerna. Sedan ger vi grundidén bakom spinnskum modellerna
i delsektion \ref{ide} och går igenom den, historiskt sett, första spinnskum modellen i delsektion \ref{pzre}. I delsektion \ref{topotri} ser vi över BF-teori och dess relation till gravitation via Plebanskis konstruktion som vi tar oss an i delsektion \ref{begr}. I delsektion \ref{secteprl} går vi igenom strukturen och uppbyggnaden av de för tillfället två bästa spinnskum modellerna av gravitation och introducerar koherenta tillstånd. Sedan anger vi de resultat som fåtts via en semiklassisk analys av nyss nämnda modeller i delsektion \ref{semian}, tillsammans med en diskussion om hur man kan beräkna gravitationspropagatorn inom denna terminologi. Till slut avslutar vi artikeln i delsektion \ref{avslut} genom att nämna om pågående forskning och en stor del av den utveckling som vi inte fått utrymme för i denna korta artikel samt de mest pressande frågorna som behöver svar.

\subsection{Prelud: Regge gravitation: \label{rggra}}

\begin{figure}[h!]
\begin{center}
\includegraphics[width=6cm]{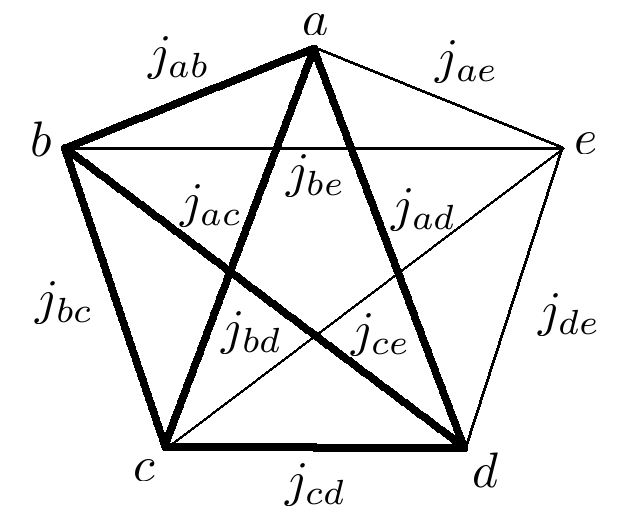}
\end{center}
\caption{Figuren visar kombinatoriken av ett 4-simplex som består av 5 st tetraedra. Tetraedern med kantlängderna $j_{ab}, j_{bc}, j_{cd}, j_{ad}, j_{ac}$ och $j_{bd}$ är
ritad med tjockare linjer för att ge en bild av hur man avläser 5 tetraedra ifrån denna grafiska representation av 4-simplexet.}
\label{foursim}
\end{figure}

Inom Regge gravitation \cite{regge} är idén att bygga upp rumtidens krökning via simpliska byggnadsblock (se t.ex. \cite{misn} för en mera detaljerad introduktion till Regge 
kalkyl eng. (\emph{Regge calculus})). Lite som att bygga upp en figur av legobitar, trots att simplexen inte har en fixerad form så som legobitar. Ett simplex är formen på byggstenarna i en 
diskretisation av en geometri. Dess struktur beror av 
dimensionen av den geometriska formen man vill approximera. För 2 dimensioner är det en triangel, för tre en tetraeder och för fyra, ett 4-simplex. 4-simplexet är en tetraeder med en femte punkt som sitter fast via sina kanter till alla fyra av tetraederns spetsar (Se Figur \ref{foursim}). 
Symmetrin är lätt att se. I två dimensioner består byggblocken av ett objekt med tre spetsar och tre kanter som sammanbinder alla dessa spetsar. I tre dimensioner, har vi fyra spetsar och därmed sex kanter som sammanbinder alla spetsar och i fyra dimensioner, fem spetsar och 10 kanter som sammanlänkar dem. D.v.s. för varje dimension adderar vi bara en ny spets och drar linjer från den till varenda annan spets, vilka vi kallar kanter. 

\begin{figure}[h!]
\begin{center}
\includegraphics[width=15cm]{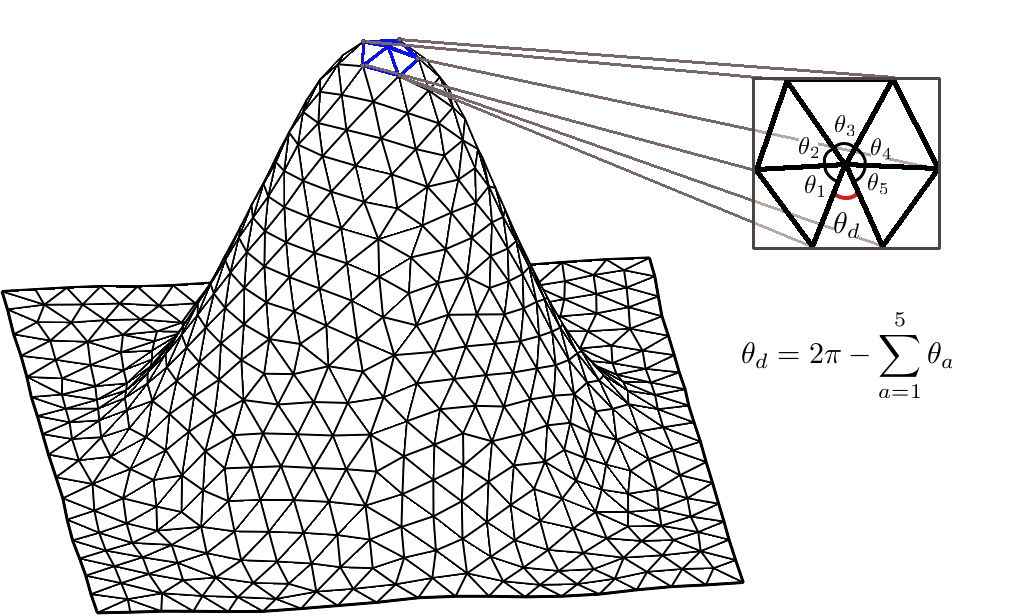}
\end{center}
\caption{Figuren illustrerar hur en två dimensionell krökt yta kan approximeras m.h.a. trianglar. I förstoringen har vi öppnat upp trianglarna 
kring en ''nod'' och projicerat dem på ett plan. Vi ser att summan av alla vinklar kring ''noden'' inte blir $2\pi$, utan är vinkeln $\sum_{a=1}^5\theta_a$ mindre
än $2\pi$. Detta är förstås en konsekvens av att geometrin på ytan är krökt.}
\label{snran}
\end{figure}

För att sammanlänka ovannämnda struktur med gravitation, och speciellt den geometriska krökning som rumtiden innefattar börjar vi med ett exempel i två dimensioner 
(se Figur \ref{snran}). 
Från Figur \ref{snran} ser vi hur vi kan approximera en två dimensionell geometri som är krökt genom att passligt placera ut platta trianglar. Då vi förstorar 
upp trianglarna kring en spets och placerar dem på en platt yta, ser vi att om vi går runt spetsen ett varv, kommer vi inte att få vinkeln $2\pi$. Detta beror på att ytan som trianglarna täckte är krökt. På detta sätt kan vi beskriva (positiv) krökning som avsaknaden av en vinkel $\theta_d$ kring spetsen som vi kallar underskottsvinkeln 
(eng. \emph{deficit angle}). M.a.o. är krökningen koncentrerad på spetsar i två dimensioner.

Underskottsvinkeln kan enkelt generaliseras till $n$-dimensioner, men dess illustrering med en figur blir mindre intuitiv ju högre 
vi går i dimension och vi nöjer oss därför med att illustrera problemet i två dimensioner. I högre dimensioner håller vi fast vid tolkningen att $n-2$
dimensionella regioner innehåller informationen om geometrins krökning (d.v.s. 0-dimensionella regioner i 2 dimensioner, spetsar).
Detta leder till att man kan generalisera tanken om den ''diskreta'' krökningen till tre dimensioner som att krökningen sitter på kanter, och i fyra dimensioner 
på areor $A_a$. Därför kan vi skriva den fyr-dimensionella approximationen av krökningen av den Riemannska geometrin $R_g$ som en summa
\be      
\sum_{a \in R_g} A_a\Theta_a, 
\ee
där summan går över alla areor $a$ inom geometrin som approximeras och $\Theta_a$ är underskottsvinkeln för arean. 
Det står klart att då vi ökar på antalet areor (block som innehåller information om geometrins krökning), kommer vi att få en bättre approximation
av den kontinuerliga geometrin. Denna idé leder till något som kallas för Regge kalkyl \cite{regge}. Inom Regge kalkyl kan man visa att integralen 
\be
{1\over 2}\int_{R_g} \sqrt{-g}R d^4x = \sum_{a \in R_g} A_a\Theta_a, 
\ee
där $g$ är determinanten av metriken och $R$ Ricci skalären, håller då trianguleringen av den fyr-dimensionella geometrin gör så fin som möjligt 
och summan är en Riemann-summa. Eftersom kvantiteten på vänster sida lätt känns igen som Einstein-Hilbert aktionen, är detta motiveringen för 
Regge kalkyl som en diskretisering av gravitation. Inom spinnskum modellerna, som vi snart skall se närmare över, använder man sig av denna typs
diskreta geometri för att placera variablerna i teorin och därför är den viktig att känna till. 

Före vi går över till spinnskum modellerna, lönar det sig ännu att 
se över terminologin för byggstenarna av simplexen, eftersom dessa förekommer i ''två'' versioner för spinnskum modellerna. En terminologi 
för konfigurationsrymden, och en för dess dual (vilket kallas för ett 2-komplex). Dualet är viktigare för spinnskum modellerna, men konfigurationsrymden, 
p.g.a. sin direkta relation till simplexen är också inom vanlig användning (se tabell \ref{onlytab} för hur terminologin relateras).
\begin{table}[h!]
\begin{center}
\begin{tabular} {c|c}
konfigurationsnamn (triangulering) & dualnamn (2-komplex) \\
\hline
spets (0) & -- \\
kant (1)& -- \\
triangel (2) & yta (2) (eng. \emph{face}) \\
tetraedra (3) & kant (1) (eng. \emph{edge}) \\
4-simplex (4) & spets (0) (eng. \emph{vertex})
\end{tabular}
\caption{Tabellen visar hur konfigurationsnamnet på de olika byggstenarna  relateras till dualnamnet i 2-komplexet. Talet inom parentes indikerar vilken dimension
byggstenen har.}
\label{onlytab}
\end{center} 
\end{table}
 Namnen för de 0 och 1 dimensionella simplexen anges också i ovanstående tabell, men dessa är inte av någon betydelse för spinnskum strukturen 
och används inte alls inom dessa modeller. Namnen för 2-komplexet är viktiga att känna igen och lära sig att identifiera med dess duala, trianguleringsnamn så att 
man har ett bättre hum om vad som diskuteras inom spinnskumretoriken. Det är att rekommendera att lära sig dessa identifikationer utantill för att lättare kunna förstå denna artikel. 

\subsection{Introduktion \label{ide}}

Då arbetet av spinnskum modellerna påbörjades \cite{rovereis, rolli}, tänkte man att spinnskum modellerna skulle representera en sorts projicering från $\mc{H}_{kin}$ på kärnan
(eng. \emph{kernel}) av Hamiltonbegränsningen så, att modellen skulle återge den fysikaliska inre produkten för SKG\footnote[1]{För en bra introduktion till dessa
idéer se \cite{ori}, trots att Barrett-Crane (BC) modellen \cite{bacr} som sedan introduceras i artikeln, visats att inte resultera i den korrekta gravitations propagatorn \cite{wpro}.}. På ett formellt plan kan detta 
anges som
\begin{align}
\langle s_i | s_f \rangle_{phys} &= \langle s_i P | s_f \rangle, \\
P &= \prod_{x \subset \Sigma}\delta(\ha{H}(x)) = \int\mc{D}[N]\exp[i\int_\Sigma \ha{N}(x)\ha{H}(x)], \label{proj}
\end{align}
där $\ha{H}(x)$ är Hamiltonfunktionen innehållande både Hamilton- och vektorbegränsningarna, $\ha{N}(x)$ förlopps och skiftes funktionerna (eng. \emph{shift function}),
$\mc{D}[N]$ måttet för stig-integralen och $\langle s_i |$ initial spinn-nätverket och $| s_f \rangle$ det slutliga spinn-nätverket. Likheten mellan produkterna av delta-funktioner i projektorn \eqref{proj} och stig integralen är endast en formell korrespondens. 

För en genomskinligare fysikalisk tolkning av \eqref{proj} är det bäst att skriva om uttrycket. Vi börjar med att separera projektorn $P$ i förlopp och 
skift delar. Då får vi 
\be
P = \int\mc{D}[N] D[g] e^{i\int_0^1dt \ha{N}(t) \ha{H}(t)}, \label{proj2}
\ee
där $D[g]$ är diffeomorfismoperatorn, $\ha{H}(t)$ är Hamiltonbegränsningen och vi har infört en parametrisering där vi har $\Sigma_i = \Sigma(0)$ 
som initial hyperyta och $\Sigma_f = \Sigma(1)$ som slutlig hyperyta. Om vi nu expanderar exponenten i \eqref{proj2}, får vi expansionen
\be
e^{i\int_0^1dt \ha{N}(t) \ha{H}(t)} = 1 +  i\int_0^1dt \ha{N}(t)\ha{H}(t) + i^2\int_0^1dt \int_0^t dt' \ha{N}(t)\ha{H}(t) \ha{N}(t')\ha{H}(t') + ... \label{crap}
\ee
Eftersom Hamiltonbegränsningen agerar genom att skapa eller förinta tre-valenta noder, kan vi dela in dess aktion på ett spinn-nätverk $| s \rangle$ 
enligt
\be
\ha{H}| s \rangle = A_v(s) D_v | s \rangle + h.c., 
\ee
där $A_v(s)$ är en amplitud, en sorts koefficienter som man kan beräkna \cite{rovereis} och $D_v$ är den delen av operatorn som förändrar på spinn-nätverket
$| s \rangle$:s struktur genom att skapa eller förinta tre-valenta noder. Med denna vetskap kan vi skriva om transitionen från ett spinn-nätverk till ett annat som 
\begin{align}
\langle s_iP | s_f \rangle &= \langle s_i| s_f \rangle +  i\big(\sum_{v \in s_i}A_v(s_i)\langle s_i| D_v | s_f \rangle \label{tramp}
+ \sum_{v \in s_f}A_v(s_f)\langle s_i| D^{\dagger}_v | s_f \rangle\big) \\ 
& + i^2 \big(\sum_{v \in s} \sum_{v \in s_1} A_v(s) A_{v'}(s_1) \langle s_i| D_v |s_1 \rangle \langle s_1|D_{v'} | s_f \rangle  +... \big) + ... \notag,
\end{align}
där vi insatt en fullständig mängd tillstånd $|s_1 \rangle \langle s_1|$ vars summa är implicit, integralerna från \eqref{crap} har absorberats in i koefficienterna $A_v(s)$ och detta gör integreringen över förloppsfunktionerna i \eqref{proj2} trivial. Integreringen över skiftesfunktionerna i \eqref{proj2} kommer att vara ekvivalent med att kräva diffeomorfisminvarians. Denna kan implementeras 
genom att t.ex. tolka spinn-nätverken i uttrycket som s-knutar eller som deras ekvivalens klasser. Huvudpoängen är att man klart ser från uttrycket \eqref{tramp} att
det är fråga om en summa av historier av spinn-nätverk (eller 3-geometrier). M.a.o. är stig-integralen för spinnskum modellerna ett sätt att ange en transition från en 3-geometri till en annan via alla spinn-nätverk som är kompatibla med dess initial och slut -tillstånd.    

Ovan nämnda geometriska tolkning är nyttig för att förstå spinnskum modellernas motivering, men i praktiken brukar de inte byggas på detta sätt. Istället brukar man utgå
från uttrycket \eqref{proj} och från en diskretisering av mångfaldet $\Sigma$ och ett bra val av variabler. Efter detta kan Peter-Weyl teoremet användas på delta-funktionerna i \eqref{proj} och dessa kan skrivas om som spåret (eng. \emph{trace}) av representationerna för gruppen i fråga (se sektion \ref{topo}).  
Produkten av delta-funktioner i \eqref{proj} delas i detta fall in i en produkt av olika delar som man kan summera i uttrycket för den fysikaliska inre produkten
\be
\langle s | s' \rangle_{phys} = \lim_{F_{s \rightarrow s'} \rightarrow \infty}\sum_{F_{s \rightarrow s'}}N(F_{s \rightarrow s'}) \sum_{\{j\}} \prod_{f \subset F_{s \rightarrow s'}}A_f(j_f) \prod_{e \subset F_{s \rightarrow s'}}A_e(j_e)\prod_{v \subset F_{s \rightarrow s'}}A_v(j_v). \label{sfamp}
\ee
Detta uttryck består av 2-komplexet $F_{s \rightarrow s'}$, som är en graf som avgränsas av spinn-nätverksfunktionerna $\langle s |$ och $|s' \rangle$, och spinn $\{ j \}$ kvanttal
angivna på 2-komplexets kanter $e$ samt dess ytor $f$. $v$ står för 2-komplexets spetsar och gränsvärdet ${F_{s \rightarrow s'} \rightarrow \infty}$ indikerar att gränsvärdet av 
oändligt många 2-komplex måste tas för att återfå alla frihetsgrader för gravitation. M.a.o. visar uttrycket 
\eqref{sfamp} att den fysikaliska inre produkten består av en summa av produkter av amplituder $A_f(j_f)$, $A_e(j_e)$ och $A_v(j_v)$ för de olika delarna av 2-komplexet. 
Detta uttryck representerar en stig-integral där det gravitationella fältet kan förstås som en historie av transitioner genom olika kvanttillstånd av rummet, enligt diskussionen ovan. Det bör poängteras att detta uttryck och spinnskum modeller kan behandla situationer där topologin förändras, vilket är en positiv utvidgning av SKG där topologin är fixerad.    

EPRL \cite{eprl} och FK \cite{fkmo} modellerna, är för tillfället de bästa kandidaterna till en spinnskum modell som teori för kvantgravitation. 
Det viktigaste med dessa modeller är att man lyckats skapa ett program m.h.a. vilket gravitationspropagatorn kan beräknas \cite{euel}. För att klara av detta har å andra sidan 
en del aspekter som funnits med från gryningen av området raderats ut. Bl.a. är inte projektionsoperatorn i EPRL och FK modellerna en riktig projektionsoperator längre eftersom $P_{EPRL/FK}^2 \neq P_{EPRL/FK}$. Dessutom konstrueras EPRL och FK modellerna genom att begränsa den topologiska BF-teorin genom Plebanskis konstruktion. Kontakten till Hamiltonbegränsningen i SKG är därför rätt lös, och man kan nog konstatera att hoppet om att de nuvarande spinnskum modellerna skulle vara en kovariant version av SKG är ganska svagt. Däremot lånar modellerna väldigt mycket från SKG som teori och tanken att en konvergens mellan dessa två teorier skulle uppnås med mera utveckling av området är inte alls omöjlig.  

I följande sektion skall vi se på spinnskum modeller i tre dimensioner och den första spinn-skum modellen, Ponzano-Regge modellen \ref{pzre}, från vilken vi lätt hoppar till 
topologisk kvantfältteori \ref {topo} som är en hörnsten i konstruktionen av EPRL och FK modellerna. Sedan ser vi på Plebanski begränsningarna i \ref{begr} som 
ger oss gravitation från den topologiska BF-teorin och i sektion \ref{secteprl} presenteras  EPRL och FK modellerna. Till slut ser vi på den semi-klassiska analysen av EPRL-modellen och beräkningen av gravitationspropagatorn i \ref{semian}.     

\subsection{3D gravitation och Ponzano-Regge modellen \label{pzre}}

I 2+1 dimensioner går gravitation att kvantisera \cite{witten}. Den första modellen i 3 dimensioner introducerades av Ponzano och Regge och kallas för 
Ponzano-Regge modellen \cite{pzmo} (se \cite{carl} för en översikt över 2+1 dimensionell gravitation). I den, börjar man från en 6j-symbol som är en viss kontraktion av fyra Clebsch-Gordan koefficienter och representerar den som 
en tetraeder. Detta fungerar eftersom 6j-symbolen satisfierar samma geometriska olikheter som kantlängderna av en tetraeder, då man associerar ett spinn-kvanttal
med varje ben i tetraedern, d.v.s. 6 stycken kvanttal. Om man sedan ser över hur 6j-symbolen beter sig asymptotiskt för stora spinn-kvanttal, får man beteendet
\be
\Big\{ \begin{array}{c c c}
j_1 & j_2 & j_3 \\
j_4 & j_5 & j_6 \label{PZa}
\end{array} \Big\} \sim {1\over \sqrt{12\pi V}}\cos(S_{Regge} + {\pi \over 4}), 
\ee   
då $V^2 > 0$, d.v.s. volymen för tetraedern är reell och en geometrisk tetraeder kan konstrueras av spinn-kvanttalen och $S_{Regge} = \sum_e(j_e + {1\over 2})\theta_e$ 
är Regge aktionen för diskret 3 dimensionell gravitation. Denna upptäckt, att $S_{Regge}$ uppstår vid gränsvärdet av stort spinn, inspirerade 
Ponzano och Regge att konstruera en kvantgravitationsmodell av 3 dimensionell gravitation med partitionsfunktionen
\be
Z_{PR} = \sum_{j_e}(-1)^\xi\prod_e(2j_e + 1)\prod_\tau\Big\{ \begin{array}{c c c}
j_1 & j_2 & j_3 \\
j_4 & j_5 & j_6 \label{PZamp1}
\end{array} \Big\}_\tau,
\ee  
där $\xi$ är en viss lineär kombination av spinn-kvanttalen, $e$ står för kanterna av tetraedern i fråga och $\tau$ för tetraedern. Som vi ser är partitionsfunktionen 
en summa av amplituder associerade till varje simplisk geometri. Likheten mellan \eqref{PZamp1} och \eqref{sfamp} är iögonenfallande. Det bör sägas att uttrycket \eqref{PZamp1} är divergent, men det går att regularisera \cite{regpz}. 

Trots att gravitation i 3 dimensioner är en topologisk teori, d.v.s. den har inga lokala frihetsgrader, är den viktig
att börja ifrån, eftersom maskineriet bakom spinnskum modellerna i 4 dimensioner bygger mycket starkt på id\'eerna i denna kvantisering. I det följande 
visar vi hur Ponzano-Regge modellen är besläktad med spinnskum modellerna (se också \cite{ale}). Vi börjar från BF-aktionen för 3 dimensionell gravitation 
\be
S(e, \omega) = \int_\mc{M}\Tr( e \wedge F(\omega)), \label{regga}
\ee
där $\mc{M} = \Sigma \times \mathbb{R}$ \footnote[1]{Notera att $\Sigma$ är 2 dimensionell i denna delsektion eftersom vi betraktar 3 dimensionell gravitation.}, $\omega$ 
är en SU(2) konnektion $F(\omega)$ dess krökning och $e$ är en $\lie{su}$(2)-värd 1-form. Gauge symmetrierna för denna aktion består av de lokala SU(2) gauge symmetrierna
\be
\delta e = [e, \alpha], \hspace{2.5 cm} \delta\omega = d_\omega\alpha, \label{inv1}
\ee
där $\alpha$ är en $\lie{su}$(2)-värd 0-form och de topologiska gauge symmetrierna
\be
\delta e = d_\omega \eta, \hspace{2.5 cm} \delta\omega = 0, \label{topog}
\ee
där $\eta$ är en $\lie{su}$(2)-värd 0-form och $d_\omega$ den kovarianta yttre derivatan (eng. \emph{exterior derivative}). Den första invariansen 
\eqref{inv1} följer från aktionen \eqref{regga} medan den andra \eqref{topog} följer ifrån Bianchi identiteten $d_\omega F(\omega) = 0$. Alla lösningar på 
rörelse-ekvationerna som följer från aktionen \eqref{regga} kommer att vara lokalt endast gauge. Detta är orsaken till att vi säger att teorin är topologisk, inga lokala 
frihetsgrader finns, endast globala eller topologiska. 

Om vi gör en 2+1 splittring av denna aktion i Ashtekarliknande variabler som vi väljer som konnektionen $A^i_a$ och $E_j^b = \epsilon^{bc}e^k_c\delta_{jk}$, 
där $a,b,c \in 1, 2$ är rumsvariabler och $i,j,k \in 1,2,3$, $\lie{su}$(2) variabler, finner vi den bekanta Poisonkommutatorn
\be
\{A^i_a, E_j^b \} = \delta^i_j\delta_a^b \delta^{(2)}(x, y)
\ee
och begränsningarna 
\begin{align}  
D_b E^b_j &= 0 \label{gss} \\
F^i_{ab} &= 0 \label{hoff}
\end{align}
Symmetrierna \eqref{inv1} och \eqref{topog} genereras av dessa begränsningar.

Eftersom spinn-nätverken automatiskt tar hand om begränsningen \eqref{gss}, återstår begränsningen \eqref{hoff}. Om vi håller oss till 
spinn-nätverks funktioner som de kinematiska tillstånden i Dirac programmet och lyckas hitta en bra definition på projektorn
\be
P =  \int\mc{D}[N]\exp[i\int_\Sigma \Tr(N \ha{F}(A))], \label{proj3d}
\ee 
har vi hittat den fysikaliska inre produkten $\langle s|s'\rangle_{phys} = \langle sP|s'  \rangle$ och klarat av Dirac programmet. Detta kan vi åstadkomma genom en 
passlig regularisering av \eqref{proj3d}. Vi diskretiserar $\Sigma$ i små två dimensionella ytor av area $\epsilon^2$, precis i enlighet 
med regulariseringen i SKG så att vi får Riemann summan
\be
P = \lim_{\epsilon\rightarrow 0}\sum_p \epsilon^2\Tr[N_p F_p(A)].
\ee 
Om vi i denna skriver $F_p(A)$ för varenda liten yta m.h.a. Wilson slingan 
\be
W_p[A] = \id + \epsilon^2F_p(A) + \mathcal{O}(\epsilon^3), \label{wloop}
\ee
och använder likheten 
\be
\int dN \exp(i\Tr[NW]) =  \sum_j (2j + 1) \Tr[D^j(W)]
\ee
som kommer ifrån Peter-Weyl teoremet, kan vi skriva om den inre produkten som
\be
\langle sP|s' \rangle = \lim_{\epsilon \rightarrow 0} \prod_p^{n_p(\epsilon)} \sum_{j_p} (2j_p + 1)\langle s \Tr[D^{j_p}(W)]| s'\rangle \label{physpro}
\ee
där spinnet $j_p$ associeras med den p:tte ytan av indelningen i små ytor av $\Sigma$ och $n_p(\epsilon)$ är antalet dessa små ytor. Gränsvärdet i 
\eqref{physpro} kan visas både existera och vara väldefinierat. 

För att introducera spinnskum representationen av denna modell inför vi en extra ofysikalisk tidsvariabel och arbetar med en indelning
av $\Sigma$ i $n_p(\epsilon)$ stycken diskreta blad (eng. \emph{foil}) $\Sigma_p$, ett för varje $\Tr[D^{j_p}(W)]$,   
och insätter en indelning av ett \footnote[1]{Denna indelning består av $\id = \underset{\gamma \subset \Sigma, \{j\}_\gamma}{\sum}|\gamma, \{j\} \rangle \langle\gamma, \{j\}|$ där summan går över alla grafer $\gamma$ och alla möjliga tilldelningar av spinn till grafens delar.},
(eng. \emph{partition of unity}) mellan varje $\Sigma_p$ så att det totala 
$\Sigma$ kan ses som en transition genom alla $\Sigma_p$ då man följer den ofysikaliska tidsvariabeln som också kan ses som modellens koordinattid. 
Detta är varför man talar om spinnskum modellerna
som en summa av historier. Vi har helt enkelt en summa av spinn-nätverk som evolverar efter en ofysikalisk tid. Noderna i 
spinn-nätverken blir kanter som innehar sammanflätare och stigarna i spinn-nätverken blir ytor som innehar spinn-kvanttal. 

Då vi gjort detta och begränsar spinn-nätverken $\langle s|$ och $|s' \rangle$ till endast 3-valenta noder återfår vi Ponzano-Regge modellens inre produkt som   
\be
\langle s| s' \rangle_{phys} = \sum_{F_{s \rightarrow s'}} \prod_{f \subset F_{s \rightarrow s'}} (2j_f +1)^{v_f\over 2}\prod_{v_f \subset F_{s \rightarrow s'}} \Big\{ \begin{array}{c c c}
j_1 & j_2 & j_3 \\
j_4 & j_5 & j_6 \label{PZamp}
\end{array} \Big\},ba
\ee
där summan $\underset{F_{s \rightarrow s'}}{\sum}$ är en summering av varje term som befinner sig på sitt egna blad $\Sigma_p$. Om vi jämför uttrycket \eqref{PZamp} med \eqref{PZamp1} ser vi att det är fråga om samma teori. M.a.o. leder kvantiseringen i spinn-skum representationen i 3 dimensioner till Ponzano-Regge modellen och är starkt 
relaterad till konstruktionen av den fysikaliska inre produkten för den kanoniska, Hamiltonska formuleringen. Frågan är hur vi skall kunna generalisera detta till en gravitation som är 4 dimensionell och har lokala frihetsgrader. Vi kommer att ta ett steg i taget och börjar med att se över kvantiseringen av $d$-dimensionell BF-teori, som också är en topologisk teori.
 
\subsection{Topologisk BF-teori \label{topotri}}

I denna delsektion ser vi över relationen mellan BF-teori och spinnskum modellerna (se också \cite{baez23}). 
I BF-teori startar vi från en kompakt grupp $G$ med Lie-algebran $g$ med en invariant inre produkt $\langle \rangle$ på det $d$-dimensionella mångfaldet 
$\mc{M}$ och aktionen 
\be
S(e, \omega) = \int_\mc{M} \langle B \wedge F(\omega)\rangle, \label{sbf}
\ee
där $B$ är en $\lie{g}$ värd $(d-2)$-form, $\omega$ en $G$-konnektion över principal knippet (eng. \emph{principal bundle}) över $\mc{M}$ och $F(\omega)$ dess krökning (jfr. med \eqref{regga}). Gauge symmetrierna för denna aktion är
\begin{align} 
\delta B &= [B, \alpha], \hspace{2cm} \delta\omega = d_\omega\alpha \label{gua} \\
\delta B &= d_\omega\eta,  \hspace{2cm} \delta\omega = 0, \label{topo}
\end{align}
där $\alpha$ och $\eta$ är $\lie{g}$ värda 0 former och $d_\omega$ den kovarianta yttre derivatan. \eqref{gua} är de lokala $G$ transformationerna och
\eqref{topo} är de topologiska gauge symmetrierna (jfr. med \eqref{inv1} och \eqref{topog}). De lokala gauge transformationerna är en konsekvens av aktionen 
medan de topologiska gauge symmetrierna är en konsekvens av Bianchi identiteten $d_\omega F(\omega) = 0$. Lösningarna på rörelse-ekvationerna 
för \eqref{sbf} kommer att vara endast gauge och därför sägs det att teorin är topologisk, den har endast topologiska frihetsgrader. 

Efter denna korta presentation av BF-aktionen, kan vi attackera partitionsfunktionen för BF-teori. Om vi för stunden antar att 
$\mc{M}$ är kompakt och orienterbart, kan vi ange den via en formell integrering över $B$ formerna som
\be
\mc{Z}_{BF} = \int \mc{D}[B]\mc{D}[\omega] \exp(i\int_{\mc{M}}\langle B \wedge F(\omega)\rangle ) = \int \mc{D}[\omega] \delta(F(\omega)). \label{zbf}
\ee
Från ovanstående uttryck kan vi avläsa att BF-partitionsfunktionen består av ''volymen'' av platta konnektioner på mångfaldet $\mc{M}$. Det formella 
uttrycket \eqref{zbf} kan definieras via en diskretisering av mångfaldet $\mc{M}$. Vi delar in det $d$-dimensionella mångfaldet cellulärt i en godtycklig cellulär
upplösning (eng. \emph{cellular decomposition}) $\Delta$.  För att dra nytta av detta, behöver vi också dess dual $\Delta^*$. Det duala cellulära 
komplexet $\Delta^*$ är ett kombinatoriskt objekt som består av spetsar $v \in \Delta^*$ duala till $d$-celler i $\Delta$, kanter $e \in \Delta^*$ duala till 
$d-1$ celler i $\Delta$ och ytor $f \in \Delta^*$ duala till $d-2$ celler i $\Delta$. Detta illustreras klart i figurerna 1-3 i \cite{alliv} då $\Delta$ är en simplisk upplösning 
i 2-4 dimensioner. Dessa celler kan vi sedan placera BF-teorins variabler på\footnote[1]{För en realistisk modell för gravitation, måste vi också se efter
den yttre gränsen av spinnskummet som BF-teorin konstrueras på och ytterligare välja variablerna på denna. Vi kommer att diskutera detta för gravitation i följande delsektion 
\ref{begr}, men för tillfället är denna komplikation onödig för vår diskussion av BF-teori.}. 

Logiken bakom valet av vilka variabler placeras på vilken cell, kommer ursprungligen ifrån den 
lyckade kvantiseringen av gravitation i 3 dimensioner. Eftersom denna kan göras på olika sätt, som leder till samma uttryck för partitionsfunktionen, kopierar man 
denna konstruktion till högre dimensioner. I 3 dimensionell gravitation i spinnskum representationen diskretiserar man konnektionen på spinnskummets kanter
och $B$-fältet (denna kallades för $e$-fältet i sektion \ref{pzre}) på dess ytor. I kvantiseringen av BF-teori gör vi på samma sätt. 

$B$-fältet associeras med Lie-algebra element $B_f$ som placeras på spinnskummets ytor. Vi kan ange $B_f$-fältet som en integral över $d-2$ cellen
som $B$-fältet är dualt till, enligt
\be
B_f = \int_{d-2 \mr{\, cellen}} B. \label{bf}
\ee 
Konnektionen $\omega$ sköter vi genom att placera ut gruppelement $g_e$ av den på kanterna av $\Delta^*$. Man kan se dessa gruppelement 
som holonomin av $\omega$ längs kanten $e$ d.v.s. 
\be
g_e = P \int_e \exp(-\omega). \label{ge}
\ee 
Användande \eqref{bf} och \eqref{ge} kan vi skriva partitionsfunktionen \eqref{zbf} som
\be
\mc{Z}_{BF}(\Delta) = \int \prod_{e \in \Delta^*} dg_e \prod_{f \in \Delta^*} dB_f e^{iB_f U_f} = \int \prod_{e \in \Delta^*} dg_e \prod_{f \in \Delta^*} \delta(g_{e_1}\cdots g_{e_n}), \label{zbf2}
\ee
där $U_f = g_{e_1}\cdots g_{e_n}$ är holonomin på kanterna som omringar en yta $f$. Vi kan nu klart se analogin till deltafunktionen i uttrycket \eqref{zbf2}. Integrationsmåttet
för $dB_f$ är det vanliga Lebesguemåttet medan måttet $dg_e$ är Haar måttet för den kompakta gruppen $G$. Deltafunktionen i \eqref{zbf2} kan skrivas
m.h.a. Peter-Weyl teoremet som 
\be
\delta(g) = \sum_\rho d_\rho \Tr[\rho(g)], 
\ee
 där $d_\rho$ är dimensionen av den unitära irreducerbara representationen av $G$ i fråga. Med denna tillsats tar uttrycket \eqref{zbf2} formen
 \be
 \mc{Z}_{BF}(\Delta) = \sum_{\mc{C}: \{ \rho \} \rightarrow \{ f \}}\int \prod_{e \in \Delta^*} dg_e \prod_{f \in \Delta^*} d_{\rho_f} \Tr [ \rho_f(g_{e_1}\cdots  g_{e_n})]. \label{zbf3}
 \ee 
Detta uttryck kan ännu integreras över gruppelementen $g_e$ då vi kommer ihåg Haar måttets egenskaper
\be
\int F(g)dg = \int F(g^{-1})dg = \int F(g h)dg = \int F(h g)dg,
\ee
där $g$ och $h$ är två gruppelement. Allt vi behöver notera är att i en simplisk upplösning $\Delta$ av det $d$-dimensionella mångfaldet $\mc{M}$, kommer kanterna i dualet
$\Delta^*$ att begränsa exakt $d$ stycken ytor $f$. M.a.o. kommer integreringen i \eqref{zbf3} att indelas i $d$ st olika spår. Med denna vetskap, kan vi använda följande 
notation 
\be
P^e_{inv}(\rho_1, ..., \rho_d) = \int dg_e \rho_1(g_e) \otimes \rho_2(g_e) \otimes \cdots \otimes \rho_d(g_e) 
\ee
för att skriva om \eqref{zbf3} som  
\be
\mc{Z}_{BF}(\Delta) = \sum_{\mc{C}_f: \{ \rho_f \} \rightarrow \{ f \}} \prod_{f \in \Delta^*} d_{\rho_f} \prod_{e \in \Delta^*}P^e_{inv}(\rho_1, ..., \rho_d). \label{zbf4}
\ee
Detta är vårt slutliga svar på partitionsfunktionen för BF-teori i d-dimensioner. Det är värt att nämna att $P^e_{inv}$ kommer att fungera som en projektor 
$P^e_{inv} = ({P^e_{inv}})^2$ till rummet $\mr{Inv}[\rho_1\otimes \cdots \otimes \rho_d]$ av invarianta tensorer i den givna representationen (jfr. med 
sammanflätare i sektion \ref{spnw}).

\subsubsection{SU(2) BF-teori i 4d \label{4dbfsu2}}

Eftersom vårt primära intresse är 4 dimensioner och EPRL och FK modellerna anger vi här hur partitionsfunktionen för BF-teori i 4 dimensioner över gruppen
$G$ = SU(2) kommer till. Startpunkten är uttrycket \eqref{zbf2} i fyra dimensioner och vi delar in hela partitionsfunktionen via en simplisk upplösning av mångfaldet $\mc{M}$ i 
4-simplex. Detta betyder att byggstenarna för vår spinnskum modell är 4-simplex, eller spetsar i den duala spinnskumsrepresentationen. Vi har alltså 
\be
\mc{Z}^{4d}_{BF} = \sum_{\mc{C}} \int \prod_{e_i = 1}^{5} dg_{e_i} \prod_{f _i = 1}^{10} \delta(h_{f_i}), 
\ee
där $e_i$ betecknar en av de fem kanterna i 2-komplexet av 4-simplexet, $h_{f_i}$ betecknar holonomin kring ytan $f_i$ och summan över $\mc{C}$ betecknar alla 
sätt att limma ihop 4-simplex för att åstadkomma trianguleringen $\Delta$ som är den simpliska uppdelningen av mångfaldet $\mc{M}$. 

För spinnskumsstrukturen är 
det viktigt att skilja mellan den yttre gränsen av spinn\-skummet och det som finns inuti den. På den yttre gränsen placerar vi gränsdatan, den som anger vilken 3-geometri 
(spinn-nätverk) vi startar och avslutar transitionen på. Därför skall denna gränsdata (eng. \emph{boundary data}) inte integreras bort, utan lämnas fri. Det betyder att då vi drar en yttre gräns kring dualet av ett 4-simplex, kommer denna nödvändigtvis att dela de ytorna $f$, som väljs att vara på den yttre gränsen, på två ställen och holonomin kring dessa gränsytor, kommer att vara ''avklippt''. Detta leder till att gaugeinvariansen för gruppen G i modellen (i vårt fall för gruppen SU(2)) kommer att gå förlorad.
För att åtgärda detta, sätter vi in en kant med ett fritt gruppelement av gruppen G mellan de två punkterna där holonomin klipps av. Denna kant bär på ett spinn och detta är spinnet på stigen för spinn-nätverket på den yttre gränsen av spinn-skummet, och är således en del av gränsdatan. På de punkter där spinnskummets kanter korsas av den yttre gränsen, placerar vi SU(2) sammanflätare. På detta sätt består gränsdatan av SU(2) spinn-nätverk, eller 3-geometrier. 

Rent praktiskt görs detta genom att indela varje kant som leder till
gränsdata i två kanter som båda integreras över. Haar måttets egenskaper tillåter detta, och den nya kanten placeras mellan de nya punkterna som kom till då de gamla kanterna delades i två. Konsekvensen av denna procedur är att partitionsfunktionen för 4d BF-teori tar formen   
\begin{align}
\mc{Z}^{4d}_{BF}(h_{f_i}) &= \sum_{\mc{C}} \int \prod_{e_iv = 1}^{10} dg_{e_iv} \prod_{f _i}^{10} \delta(g_{e_iv}h_{f_i}g_{e_jv^{-1}}), \label{4s}
\end{align}
där vi indelat varje kant i två, så att vi får 5 nya integraler över gruppelement och partitionsfunktionen är tydligt en funktion av gränsdatan $h_{f_i}$. Notationen 
$e_iv$ anger att kanten $e_iv$ är orienterad i riktning från spetsen $v$ till $e_i$, medan $e_jv^{-1}$ är orienterad från $e_j$ till $v$. Uttrycket \eqref{4s} gäller gränsdatan för ett ensamstående 4-simplex, i en mera generisk kalkyl
måste vi först välja gränsdatan (spinn-nätverken) och sedan anpassa byggstenarna (4-simplexen) till denna data. Då kan vi få situationer där färre, eller flere,
än 10 ytor ''avklipps'' av den yttre gränsen. Det viktiga är att byggstenarna är 4-simplex som kommer med en viss amplitud, spetsamplituden (eng. \emph{vertex amplitude}), 
och denna amplitud beror inte på hur gränsdatan är arrangerad, utan endast på hur många 4-simplex eller spetsar vi har.  

Vi kan fortsätta på vår kalkyl genom att använda Peter-Weyl teoremet för gruppen SU(2) och skriva varje $\delta(g)$ funktion som ett 
spår över unitära irreducerbara representationer av gruppen SU(2). I vårt fall $\delta(g) = \sum_j d_j\Tr [D^j(g)]$,
där $d_j = 2j+1$ är dimensionen för representationen $D^j(g)$. Då vi ytterligare vet att varje kant i dualet av 4-simplexet representerar en 
tetraeder, som har fyra trianglar, duala till ytor, kommer varje integration över gruppelementen på kanterna att fördelas på tensorprodukten av fyra spår över gruppelement.
Eftersom det finns fem kanter i dualet av 4-simplexet, kommer vi att ha 5 gruppintegreringar att göra\footnote[1]{Eftersom vi delade upp alla 5 kanter i två och skapade 10 integraler över gruppelement i \eqref{4s}, kan man undra vart de 5 andra integralerna tog vägen. De hör ihop med gränsdatan och absorberas av denna. De är givetvis viktiga när vi känner 
till gränsdatan och måste väljas passligt i det fallet, men eftersom vi nu endast är intresserade av spetsamplituden för att kunna ge ett generellt uttryck för spinnskum modellens 
partitionsfunktion, väljer vi att ignorera dessa integraler. De är i princip nu en del av argumentet i $h_{f_i}$ i partitionsfunktionen.}
över 4 spår av representationer av gruppelement. Integralerna kan beräknas enligt
\begin{align}
\int_{\mr{SU}(2)} dg D^{j_1}_{m_1m'_1}(g) \otimes D^{j_2}_{m_2m'_2}(g) \otimes D^{j_3}_{m_3m'_3}(g) \otimes D^{j_4}_{m_4m'_4}&(g) =  
  \sum_i \bar{\mathtt{i}}_{m_1m_2m_3m_4}\mathtt{i}_{m'_1m'_2m'_3m'_4}, \label{suint} \\
\mathtt{i}_{m_1m_2m_3m_4} = \left(\begin{array}{c c c} j_1 & j_2 & i \\ 
m_1 & m_2 & m \end{array}\right)g^{mm'}_i \left(\begin{array}{c c c} i & j_3 & j_4 \\ 
m' & m_3 & m_4 \end{array}\right), \hspace{-3cm} &  
\end{align}  
där $g^{mm'}_i$ är Killing-Cartan metriken för gruppen G, som i vårt fall av SU(2) endast är Kronecker delta-symbolen för spinn $j$, $\delta^{mm'}_j$. Då vi kombinerar spåren av de fem integralerna över SU(2) i \eqref{4s} enligt \eqref{suint}, får vi efter lite kombinatorik uttrycket 
\be
\mc{Z}_{BF}^{4d}(h_{f_i}) = \sum_\mc{C} \prod_{f} d_{f} \prod_{v} \{15j\}_{v}, \label{crappp}
\ee
där $v$ står för spinnskummets spetsar och $\{15 j\}$ är en 15-j symbol, som är en kontrahering av Clepsh-Gordan koefficienter, precis
som 6j-symbolen i sektion \ref{pzre}. Denna modell, men dubblerad\footnote[1]{Med detta menas att vi har två okopplade kopior av \eqref{crappp} stig-integralen.} är 
relevant för den Euklidiska gravitationen som konstrueras på $G$ = SU(2) $\times$ SU(2), för gravitation. Då vi har en Lorentzisk signatur använder vi 
gruppen $G$ = SL$(2, \mathbb{C})$. Gruppen  SL$(2, \mathbb{C})$ är inte kompakt, men detta problem kan lätt åtgärdas genom att fälla bort en av 
integreringarna över gruppen i stig-integralen, vilket motsvarar att fastställa gaugen för gruppen partiellt \cite{regeprl}. 

\subsection{Plebanski begränsningarna \label{begr}}

För att komma åt gravitation via topologisk BF-teori måste man införa begränsningar så, att krökningen $F(\omega)$ inte är platt. Detta betyder att man genom 
begränsningar frigör lokala frihetsgrader så att teorin inte är topologisk längre. För fyra dimensionell BF-teori, kan detta göras via Plebanskis \cite{pleb}\footnote[2]{Egentligen är 
aktionen som används i spinnskum modeller en modifierad Plebanski aktion, därför användningen av notationen $\ti B$, och inte den ursprungliga använd av 
Plebanski som bygger på $B$-fältet i BF-teori, men förändringen är inte stor. Därför kommer 
vi att fortsätta att hänvisa till den som Plebanskis aktion i denna artikel.} (se också \cite{kiri}) formulering 
av gravitation som startar från aktionen
\be
S = \int \ti B^{IJ}\wedge F_{IJ}(\omega) - {1\over 2}\phi_{IJKL}\ti B^{IJ}\wedge \ti B^{KL} + \mu H(\phi), \label{pleb}
\ee 
där $\ti B^{IJ}$ är en $\lie{g}$ värd 2-form, $\phi_{IJKL} = -\phi_{JIKL} = -\phi_{IJLK} = \phi_{KLIJ}$ och $\mu$ Lagrangemultiplikander, $H(\phi) = a_1\phi_{IJ}^{\, \, \, \, \, \, \, IJ} + a_2\phi_{IJKL}\epsilon^{IJKL}$ och $a_1$ och $a_2$ två godtyckliga konstanter. Variation i avseende å Lagrangemultiplikanden $\mu$ ger oss begränsningen
$H(\phi) = 0$ och variation i avseende å multiplikanden $\phi$ tillsammans med begränsningen $H(\phi) = 0$ ger oss begränsningarna på $\ti B$-fältet. Dessa är
\begin{align}
\ti B^{IJ} \wedge \ti B^{KL} &= {1\over 6}(\ti B^{MN}\wedge \ti B_{MN})\eta^{[I|K|}\eta^{J]L} - {1\over 12}(\ti B^{MN}\wedge \star \ti B_{MN})\epsilon^{IJKL} \label{Beqn} \\
0 &= \epsilon a_1\ti B^{IJ}\wedge \star \ti B_{IJ} - 2a_2\ti B^{IJ}\wedge \ti B_{IJ}, \label{epseq}
\end{align}  
där $\star$ är Hodge operatorn som på 2-formen $\ti B$ agerar enligt $\star \ti B_{IJ} = {1\over 2}\epsilon_{IJKL} \ti B^{KL}$ och $\star^2 = \pm 1$ beroende på vilken 
signatur mångfaldet har. $+1$ för Euklidisk och $-1$ för Lorentzisk. $\epsilon$ parametern i \eqref{epseq} beror på samma sätt av signaturen. Den är $+1$ 
för Euklidisk och $-1$ för Lorenzisk signatur. Dessa begränsningar kan visas vara ekvivalenta \cite{mer4, liva} till begränsningen 
\be
\ti B = \alpha \star e \wedge e + \beta e \wedge e, \hspace{2cm} {a_2\over a_1} = {\alpha^2 - \beta^2\over 2\alpha\beta}, \label{bfcon}
\ee
där $e$ är en tetrad. Detta visar klart och tydligt begränsningarnas betydelse. Med dessa val har vi kunnat begränsa $\ti B$-fältet så att vi återfår Holst-aktionen
\eqref{holst} med Barbero-Immirzi parametern $\gamma = {\alpha\over \beta}$, aktionen för SKG. Vi kan därför med gott samvete påstå att gravitation är en topologisk BF-teori med begränsningar. Trots detta finns det en liten 
sak att vara försiktig med i denna formulering. Teorin är nämligen invariant under transformationen $\ti B \rightarrow \star \ti B$. Denna transformation byter om rollerna för konstanterna 
$\alpha$ och $\beta$ och inverterar därför parametern $\gamma$. Detta betyder att vi genom användning av Hodge operatorn på $\ti B$-fältet kommer att hitta 
fyra olika sektorer för gravitation. De kan anges, efter successiv användning av Hodge operatorn, som
\be
(\alpha, \beta) \rightarrow (\beta, -\alpha) \rightarrow (-\alpha, -\beta) \rightarrow (-\beta, \alpha), 
\ee   
sektorerna. Dessa sektorer måste separeras då man konstruerar spinnskum-amplituder. Oftast tas detta problem om hand i konstruktionen av spets-amplituden 
för spinnskummet. Orsaken man måste beakta denna separering härstammar ifrån en betraktelse av vanlig allmän relativitet från Plebanski 
aktionen. Om vi sätter $a_1 = 0$, i begränsningen \eqref{bfcon} får vi de fyra sektorerna
\begin{align}
& (I \pm)\, \, \,\,\,  \ti B = \pm \star e\wedge e \label{I}\\ 
& (II \pm)\,\,  \ti B = \pm e\wedge e \label{II},
\end{align} 
där sektorerna $(II \pm)$ inte är allmän relativitet, utan en icke-geometrisk teori med torsion. Därför måste sektorerna $(II \pm)$ kunna hittas ur spinnskum amplituderna
så att man kan göra sig av med dem i den slutliga modellen. Dessa sektorer sammanfaller då tetraden $e$ är degenererad. M.a.o. kan man navigera mellan dessa olika 
sektorer via de degenererade konfigurationerna i stig-integralen.  

För att ange begränsningen \eqref{bfcon} på $\ti B$-fältet i en form som är lättare att implementera i spinnskum modellerna, kan vi skriva om den m.h.a. $B$-fältet 
i vanlig BF-teori som
\begin{align}
\epsilon_{IJKL}B^{IJ}_{\mu\nu}B^{KL}_{\rho\sigma} = e\epsilon_{\mu\nu\rho\sigma}, \label{Mbeg}
\end{align}
där $B = \star e \wedge e$ och $e = {1\over 4!}\epsilon_{IJKL}B^{IJ}_{\mu\nu}B^{KL}_{\rho\sigma}\epsilon^{\mu\nu\rho\sigma}$. Detta är fördelaktigt 
eftersom vi från BF-teori vet att $B$-fältet skall diskretiseras på spinnskummets ytor $f$. Vi kan därför dela in dessa begränsningar enligt följande
\begin{align}
\epsilon_{IJKL} B^{IJ}_f B^{KL}_f & = 0, \hspace{1.3cm} f \in v \hspace{2cm} (\mr{Trianglar}) \label{firstC} \\
\epsilon_{IJKL} B^{IJ}_f B^{KL}_{f'} & = 0, \hspace{0.8cm} f, f' \in v \hspace{2cm} (\mr{Tetraedra}) \label{middleC} \\
\epsilon_{IJKL} B^{IJ}_f B^{KL}_{\bar{f}} & = e_v, \hspace{0.73cm} f, \bar{f} \in v \hspace{1.9cm} (4-\mr{simplex}) \label{lastC} 
\end{align} 
där parenteserna på höger sida nämner på vilka delar av diskretiseringen ifrågavarande begränsning skall gälla. Begränsningen på trianglar 
gäller alla ytor $f$ som rör spetsen $v$, begränsningen på tetraedra gäller alla ytor $f$ och $f'$ vars duala trianglar finns i samma tetraeder i 4-simplexet som definieras
av spetsen $v$ och begränsningen på 4-simplex gäller alla ytor $f$ och $\bar{f}$ vars duala trianglar delar på endast en punkt. Denna sista begränsning \eqref{lastC} brukar 
tolkas som en definition på 4-volymen $e_v$ för 4-simplexet. Oberoende av vilka par $f$ och $\bar{f}$ vars duala trianglar delar på endast en punkt väljs, kommer 
4-volymen för 4-simplexet att vara $e_v$. P.g.a. vårt val av variabler, kommer begränsningen \eqref{lastC} att gälla då alla andra begränsningar \eqref{firstC}, \eqref{middleC} 
och $\underset{f\in t}{\sum} B^{IJ}_f = 0$\footnote[1]{Denna begränsning gäller i kvantteorin eftersom det är den som står för den interna gaugen, vilken spetsamplituden
i modellerna vi konstruerar kommer att projicera på. I princip är det fråga om Gauss-begränsningen som användningen av spinn-nätverk sköter om.}, där $t$ står för 
tetraeder gäller. M.a.o. måste vi begränsa teorin med \eqref{firstC} och \eqref{middleC} för att komma åt gravitation. 

För att göra detta generaliserar vi \eqref{middleC} till
\be
n_I(\star B_f^{IJ}) = 0, \label{ncon}
\ee    
där $n_I$ en intern vektor för vilken ovanstående uttryck gäller för alla $f \in t$. Denna begränsning är nu starkare än \eqref{middleC}, eftersom 
den endast väljer sektorn $B = \pm\star e \wedge e$. Sedan diskretiserar vi Holst aktionen på ytor $f$, så att vi får uttrycket
\begin{align}
S(B_f, U_f) &= {1\over 2\kappa}\sum_{f \in \Delta^*} \Tr [B_f(t) U_f(t,t) + {1\over \gamma}\star B_f(t)U_f(t,t)]  \label{diska} \\  
&+ {1\over 2\kappa}\sum_{f \in \partial\Delta^*} \Tr [B_f(t) U_f(t,t') + {1\over \gamma}\star B_f(t)U_f(t,t')], \notag
\end{align}
där $\kappa = 8\pi G$, $U(t,t')$ är holonomin kring en yta $f$ som börjar i tetraeder $t$ och slutar i $t'$ (kom ihåg att tetraedrarna är duala till kantar och de begränsar ytan $f$ i 
dualet $\Delta^*$) och den första summan gäller endast innandömet av 
den simpliska upplösningen $\Delta^*$ av spinnskummet medan den andra summan endast går över den simpliska upplösningens yttre gräns $\partial\Delta^*$\footnote[2]{Denna
användning av $\Delta^*$ gäller endast ekvation \eqref{diska} i denna presentation. Hädanefter fortsätter $\Delta^*$ att bemärka den totala simpliska upplösningen, både 
innandöme och yttre gräns}. Aktionen \eqref{diska} är i princip Holst aktionen diskretiserad och given i BF variabler, där F nu blev holonomin $U$ p.g.a. användningen av 
\eqref{wloop}. 

Den konjugata variabeln till $U(t,t')$ på spinnskummets yttre gräns är det höger-invarianta vektorfältet för Lie gruppen $\lie{g}$. Vi kallar den för $J_f(t)$ och 
den anges av
\be
J_f(t) = {1\over \kappa}(B_f + {1\over \gamma} \star B_f).
\ee
Detta vektorfält är också det 
Hamiltonska vektorfältet för samma grupp. M.a.o. kommer den konjugata variabeln till $U(t,t')$ att representera höger-invarianta vektorfält som bestäms 
av basen i $\lie{g}$ och vi kommer därför att kunna identifiera dem med generatorerna för gruppen ifråga. Detta är av betydelse i kvantteorin och därför anger vi 
nu begränsningarna \eqref{middleC} och \eqref{ncon} i dessa variabler i Lorentzisk signatur som
\begin{align}
\star J_f \cdot J_f(1 - {1\over \gamma^2}) - {2\over \gamma} J_f\cdot J_f &= 0 \label{const11} \\
n^I(\star J_f + {1\over \gamma} J_f) &= 0. \label{const22}
\end{align}
Detta är vår slutliga form på de begränsningar som vi kommer att måsta promovera till begränsningar i kvantteorin och begränsa BF-amplituderna med för att
få en teori om gravitation. 

Spinn-skum id\'een torde nu vara klar. Vi tar den 4-dimensionella BF-teorin med gruppen G = SL($2, \mathbb{C}$) och begränsar dess amplituder på ett 
passligt sätt via Plebanski begränsningarna så att vi kan bygga upp en spinnskum modell i 4-dimensioner för gravitation. Detta gör vi i nästa sektion.
 
\subsection{Lorentziska spinnskum modeller \label{secteprl}}

För att konstruera EPRL amplituden \cite{eprl, epr, epr2} från BF-teori och uttrycket \eqref{zbf2} behöver vi först Peter-Weyl representationen för delta funktionen för
gruppen $G = $SL($2, \mathbb{C}$)\footnote[1]{För en översikt över representationsteorin för SL($2, \mathbb{C}$) se t.ex. \cite{sl2c}}, vilken är
\be
\delta(g) = \sum_k \int_{\mathbb{R}^+} dp (p^2 + k^2) \sum_{j,m} D^{p,k}_{jm jm}(g).  \label{deltaf}
\ee
där $j \geq k$ och $j \geq m \geq -j$. Vi observerar att SL($2, \mathbb{C}$) representationerna bemärks av två parametrar $p$ och $n = 2k$,
där $p \in \mathbb{R}$ och $n \in {\mathbb{Z}^+}$\footnote[2]{SL($2, \mathbb{C}$) bemärks egentligen av parametrarna $p$ och $n$, inte $k$. Men 
eftersom begränsningen \eqref{const22} kommer att begränsa oss till representationer med parametern $k = {n\over 2}$ är representationen av delta funktionen 
\eqref{deltaf} den vi är intresserade av.}. Nästa steg är att välja en godtycklig SU(2) undergupp i gruppen 
SL($2, \mathbb{C}$). Detta gör vi för att kunna projicera SL($2, \mathbb{C}$) representationerna för den yttre gränsen av spinnskummet 
på SU(2)\footnote[3]{Egentligen får vi sk. projicerade spinn-nätverk \cite{prosn} men dessa är så gott som samma sak som SU(2) spinn-nätverken.}. På detta sätt skapar vi SU(2) spinn-nätverk på den yttre gränsen av spinnskummet, vilket är hela idéen med spinnskum modellerna: en stig-integral som
tar oss från ett SU(2) spinn-nätverk (en 3-geometri) till ett annat.    

Då vi väljer denna undergrupp kommer vi att behöva uppdela representationsrummet för Hilbertrummet $\mc{H}_{p,k}$ för varje SL($2, \mathbb{C}$) representation, 
i representationsrummet $\mc{H}^j_{p, k}$ för de unitära irreducerbara representationerna för SU(2), $D^j$. Detta kan vi göra enligt följande 
\be
\mc{H}_{p,k} = \bigoplus_{j = k}^\infty \mc{H}^j_{p,k}. \label{decompH}
\ee 
Undergruppen SU(2) kommer att spela en stor roll också i valet av begränsningar. Vi väljer en tidslik vektor 
$n^I = (1,0,0,0)$ i \eqref{const22} för att kunna använda en teknik som vi snart kommer att introducera för att ålägga begränsningarna på spinnskummet. Denna 
tidslika vektor hålls invariant under aktionen av den godtyckliga SU(2) undergruppen. Valet av vektorn $n^I$ ger oss begränsningen 
\be
{1\over 2}\epsilon^i_{jk}J^{kl}_f + {1\over \gamma}J^{i0}_f = L^i_f + {1\over \gamma} K^i_f = 0. \label{qcon}
\ee
med en godtycklig SU(2) undergrupp i SL($2, \mathbb{C}$), där $L^i_f$ är generatorn för rotationer och $K^i_f$ generatorn för lyft (eng. \emph{boosts}). 
Det resulterar också i ett partiellt val av gauge i denna modell. Problemet med begränsningen \eqref{qcon} är att dess kommutator inte sluter sig, vilket resulterar 
i att vi inte kan ålägga begränsningen \eqref{qcon} starkt, som en operatorekvation. Dessutom är \eqref{qcon} inte gaugeinvariant. Men, eftersom begränsningen 
\eqref{qcon} är ekvivalent rent klassiskt, till begränsningen
\be
M_f = (L^i_f + {1\over \gamma} K^i_f)^2 = 0, \label{master}
\ee 
kan vi använda oss av denna. Begränsningen \eqref{master} är gaugeinvariant och dessutom positiv definit och därför ekvivalent till \eqref{qcon}. 
Resultatet är att EPRL modellen i denna form endast gäller rums-lika hyperytor, men konstruktionen kan också generaliseras att gälla tidslika hyperytor
\cite{jeff1, jeff2, jeff3}.

För att få de slutliga begränsningarna i kvantteorin, kan vi beräkna \eqref{master} och sedan använda 
begränsningen \eqref{qcon1}, som är begränsningen \eqref{const11} i kvantkläder, för att få de två slutliga begränsningarna 
\begin{align}
C_2(1 - {1\over \gamma^2}) + {2\over \gamma}C_1 &= 0 \label{qcon1} \\
C_2 - 4\gamma L^2 &= 0 \label{qcon2}
\end{align}
där $C_2 = \star J \cdot J = -4L\cdot K$ och $C_1 = J\cdot J = 2(L^2 + K^2)$ är Casimiroperatorerna för SL($2, \mathbb{C}$). Dessa Casimiroperatorer 
anges av
\begin{align}
C_1 &= {1\over 2}(n^2 - p^2 - 4) \\
C_2 &= np.
\end{align}
Då vi sätter in dessa värden i \eqref{qcon1} får vi
\be
np(\gamma - {1\over \gamma}) = p^2 - n^2, 
\ee
med lösningarna $p = n\gamma$ och $p = -{1\over \gamma}n$. Vi kommer att strunta i den senare lösningen eftersom den reflekterar 
sektorn $B = e\wedge e$ som begränsning \eqref{qcon2} gör oss av med (se sektion \ref{begr}). Begränsningen \eqref{qcon2} ålägger 
också kravet $n = 2k$ och vi kommer hädanefter att endast använda parametern $k$ för bemärkning av denna parameter i representationerna
av SL($2, \mathbb{C}$). Vi märker också att representationerna av lägsta vikt $j = k = {n\over 2}$ i uppdelningen \eqref{decompH} är de som kommer att 
spela en roll i denna modell. 
 
Sättet som vi ålägger begränsningen \eqref{master} är värt att notera. Det är i princip ekvivalent med Gupta-Bleuler metoden att 
ålägga begränsningar svagt enligt 
\be
\langle \Psi | L^i_f + {1\over \gamma} K^i_f) |\Psi ' \rangle =  0, \label{GBcon}
\ee
där $\langle \Psi|$ och $|\Psi '\rangle$ är SL($2, \mathbb{C}$) funktioner av formen 
\be
\Psi(g) = \sum_{j, mm'} c_{j, mm'} D^{\gamma j, j}_{jm, jm'}(g)
\ee
och $c_{j, mm'}$ konstanter. Om man ålägger begränsningarna enligt \eqref{GBcon}, 
resulterar detta i kravet att representationerna för SL($2, \mathbb{C}$) begränsas av $p = \gamma(j +1)$ och $k = j = {n\over 2}$. Detta 
alternativ för val av $p$ är dåligt p.g.a. att det ger $j = 0$ tillstånd en icke-trivial vikt i stig-integralen och detta är i stark konflikt med 
diffemorfisminvarians. Det leder också till en ny frihetsgrad i teorin, som är i konflikt med antalet frihetsgrader i fasrummet för SKG \cite{ha}. Därför valet 
$p = \gamma j$ i EPRL-modellen. Man kan karakterisera detta val genom att säga att relationen \eqref{GBcon} håller vid det semi-klassiska gränsvärdet, $\hbar \rightarrow 0, j \rightarrow \infty, \hbar j = \mr{konst.}$, av EPRL-modellen.

\subsubsection{EPRL-modellen}

Nu har vi äntligen allt vi behöver för att bygga upp transitionsamplituderna i EPRL modellen. Vi börjar från uttrycket \eqref{zbf2} och använder det 
för att skriva uttrycket för amplituden för ett ensamstående 4-simplex i BF-teori (jfr. med sektion \ref{4dbfsu2}). Detta ger oss
\begin{align}
A_{BF}^{4-S}(h_{ij}) = \int \prod_i dg_i \prod_{i < j} \delta(g_i h_{ij} g_j^{-1}), \label{bfpart}
\end{align}
där holonomin runt en yta i $\Delta^*$ anges av $U_f = g_i h_{ij} g_j^{-1}$. Sedan använder vi \eqref{deltaf} för att ange amplituden för 4-simplexet m.h.a.
representationer av SL($2, \mathbb{C}$), som vi slutligen integrerar användande följande uttryck
\begin{align}
\int_{\mr{SL(}2, \mathbb{C}\mr{)}} dg & D^{p_1, n_1}_{j_1m_1j'_1m'_1}(g) \otimes D^{p_2, n_2}_{j_2m_2j'_2m'_2}(g)  \otimes
\bar{D}^{p_3, n_3}_{j_3m_3j'_3m'_3}(g) \otimes \bar{D}^{p_4, n_4}_{j_4m_4j'_4m'_4}(g) = \notag \\
&\sum_{n}\int dp(n^2 + p^2)C^{n, p}_{(j_1m_1)\cdots(j_4m_4)} \bar{C}^{n, p}_{(j'_1m'_1)\cdots(j'_4m'_4)}, \label{intertw}
\end{align}
där $C^{n, p}_{(j_1m_1)\cdots(j_4m_4)}$ är sammanflätare i $n, p$ representationen av SL($2, \mathbb{C}$) och består av 
Clepsh-Gordan koefficienter av Lorentz gruppen enligt
\be
C^{n, p}_{(j_1m_1)\cdots(j_4m_4)} = \sum_{j, m}C^{n_1p_1n_2p_2np}_{(j_1m_1)(j_2m_2)(jm)}\bar{C}^{n_3p_3n_4p_4np}_{(j_3m_3)(j_4m_4)(jm)},
\ee 
där $C^{n_1p_1n_2p_2np}_{(j_1m_1)(j_2m_2)(jm)}$ är en Clepsh-Gordan koefficient. Denna gymnastik resulterar i
\be
A_{BF}^{4-S}(h_{ij}) = \sum_{n_f}\int dp_f (n^2_f + p^2_f)\sum_{n_e}\int dp_e(n_e^2 + p_e^2) 15j[(n_e, p_e):(n_f, p_f)]\Psi_{n, p}(h_{ij}),
\ee
där $\Psi_{n, p}(h_{ij}) = \bigotimes_e \bar{C}^{n_e, p_e} \bigotimes_{f} D^{n_f, p_f}(h_{ij})$ är en sorts  SL($2, \mathbb{C}$) spinn-näts funktional med 
SL($2, \mathbb{C}$) representationer på ytorna $f$ och SL($2, \mathbb{C}$) sammanflätare på kanterna $e$ och $15j[(n_f, p_f):(n_e, p_e)]$ är
en SL($2, \mathbb{C}$) 15j-symbol. Detta är 4-simplex amplituden för BF-teori, men genom att använda begränsningarna som ger oss $j = k = {n\over 2}$ och 
$p = \gamma j$ får vi 4-simplex amplituden för gravitation. Då vi tar skalärprodukten av den med ett SU(2) spinn-nätverk $\psi_{j_f, \mathtt{i}_e}(h_{ij})$ får vi vårt slutliga
uttryck för EPRL-modellens 4-simplex amplitud. Det blir   
\be
A_{EPRL}^{4-S}(j_f, \mathtt{i}_e) =  \sum_{n_e}\int dp_e(n_e^2 + p_e^2) 15j[(n_e, p_e):(2j_f, 2\gamma j_f)]\Big(\bigotimes_{e} f^{\mathtt{i}_e}_{n_e j_e}(j_f) \Big),
\ee
där $f^{\mathtt{i}}_{np} = \mathtt{i}^{m_1m_2m_3m_4}\bar{C}^{np}_{(j_1m_1)\cdots(j_4m_4)}$. När vi generaliserar detta uttryck till en kollektion 
4-simplex och limmar ihop dem enligt gränsdatan vi betraktar, får vi den slutliga partitionsfunktionen som 
\be
\mc{Z}_{EPRL} = \sum_{j_f, \mathtt{i}_e} \prod_f (2j_f)^2(1 + \gamma^2)\prod_v A_v(j_f, \mathtt{i}_e).
\ee
Denna modell har också den speciella egenskapen att arean är kvantiserad precis enligt SKG. Vi kan se detta genom att 
betrakta arean på en triangel 
\be
A^2 = {1\over 2}(\star B)^{ij}(\star B)_{ij} = \Big({\kappa\gamma^2 \over \gamma^2 + 1}\Big)^2\Big(K - {1\over \gamma} L \Big)^2.
\ee
Då vi använder begränsningarna \eqref{qcon1}  och \eqref{GBcon} får vi 
\be
A^2 = \kappa^2\gamma^2L^2 \Rightarrow A = \kappa\gamma \sqrt{j(j+1)}, 
\ee 
areaspektret för SKG.

\subsubsection{Koherenta representationer}

Det finns ett annat sätt att kvantisera en spinnskum modell med Lorentzisk signatur som kallas för FK modellen. Den använder sig av 
koherenta representationer för SU(2) som vi sätter sätter in mellan två representationer av SL($2, \mathbb{C}$) som separerar åt 
två 4-simplex \cite{simoli}. T.ex. sätter vi in 4 st såna i \eqref{intertw}. Dessa är resolutioner av identitesrepresentationen i SU(2) som minimerar 
osäkerheten i $\lie{su}(2)$s Casimiroperator $J^2$. De anges av uttrycket
\be
\id_j = \sum_m |j, m\rangle \langle j, m| = (2j + 1)\int_{\mr{SU(2)}} dg |j, g\rangle \langle j, g| \label{idres}
\ee  
där $\id_j$ är identiteselementet av SU(2) i representationen $j$ och 
\be
|j, g\rangle = g |j, j\rangle = \sum_m |j, m \rangle D^{j}_{mj} (g), 
\ee
är ett koherent tillstånd. Integralen i \eqref{idres} kan också anges som en integral över sfären $S^2 = SU(2)/U(1)$ eftersom $D^j_{mj}(g)$ och
$D^j_{mj}(hg)$ endast skiljer sig med en fas av gruppelementet $h$. Därför kan vi också ange \eqref{idres} som
\be
\id_j = (2j+1)\int_{S^2} dn|j, n \rangle \langle j, n |,
\ee  
där $n \in S^2$ integreras med det invarianta måttet för sfären $S^2$. Från dessa tillstånd är det klart att för generatorerna av $\lie{su}$(2), $J^i$, gäller
\be
\langle j,n |J^i |j, n\rangle = jn^i,
\ee
där $n^i$ är en tre dimensionell enhetsvektor på $S^2$. Vi kan också beräkna $J^2$ Casimiroperatorns fluktuationer vilka är minimala, 
d.v.s. $\Delta J^2 = \hbar^2 j$ för tillstånden $|j, n\rangle$ och därför också för de koherenta tillstånden. Dessa fluktuationer går till 0 
vid gränsvärdet $\hbar \rightarrow 0, j \rightarrow \infty, \hbar j = konst.$, som brukar kallas det semi-klassiska gränsvärdet för spinnskum
modellerna. Tillståndet $|j, n\rangle$ är ett semi-klassiskt tillstånd som beskrivs av en vektor i $\mathbb{R}^3$ med riktning $n$ och längd $j$. 

En viktig egenskap som de har, som vi kommer att använda då vi ser på den semi-klassiska analysen i delsektion \ref{semian} är följande:
\be
|j, j \rangle = |{1\over 2}, {1\over 2}\rangle \cdots  |{1\over 2}, {1\over 2}\rangle  \equiv |{1\over 2}, {1\over 2}\rangle ^{\otimes 2j},
\ee
vilken leder till
\be
|j, n \rangle = |{1\over 2}, n\rangle^{\otimes 2j} \label{koheg}.
\ee
Användningen av dessa tillstånd har en intressant fördel, förutom att de leder till FK-modellen, skapar de också möjligheten att tolka dem som 
''vågpaket'' som har sin ''topp'' på diskreta klassiska intrinsiska och extrinsiska geometrier. Detta kan vi utnyttja för att studera spinnskum modellernas 
semi-klassiska egenskaper. Vi kan bl.a. skriva transitionsamplituden $W(j_l, h_l)$ i EPRL modellen som 
\begin{align}
W(j_l, h_l) &=  \lim_{\mc{K} \rightarrow \infty}\sum_{j_f}\Big\{\big[\prod_{v, e \in \partial v}\int_{\mr{SL}(2, \mathbb{C})}dg_{ev}\big]\big[\prod_{e \in \partial f, f}\int_{S^2}dn_{ef}\big] \label{trnamp} \\
& \hspace{5.5cm} \times \prod_f\mu(j_f)\prod_{v \in f}\langle j_f, n_{ef} |Y^\dagger_\gamma g_{ev}g_{ve'} Y_\gamma | j_f, n_{e'f}\rangle\Big\}, \notag \\
Y_\gamma &:  \hspace{10pt} |j, n\rangle \rightarrow |(\gamma j, j), j, n\rangle,
\end{align}
där $| j_f, n_{e'f}\rangle$ är ett koherent tillstånd, $Y_\gamma$ är en projektion från SU(2) Hilbertrummet $\mc{H}^j$ till SL(2, $\mathbb{C}$) Hilbertrummet 
$\mc{H}^{\gamma j, j}$, $\mu(j_f)$ är vikten på varje yta, vilken brukar väljas som $\mu(j_f) = 2j_f + 1$ och varje kant har delats in i två halva kanter för att möjliggöra 
insättningen av SU(2) koherenta tillstånd. Parenteserna i \eqref{trnamp} indikerar hur produkterna delar upp sig. 

\subsubsection{FK-modellen}

I FK modellen \cite{fkmo} använder man sig av de koherenta tillstånden. Idéen bygger helt enkelt på att begränsningen \eqref{GBcon} 
kan åläggas semi-klassiskt \'a la Gupta-Bleuler direkt på de koherenta tillstånden \eqref{idres} som tar del i ett liknande uttryck som \eqref{trnamp} fast för FK-modellen. Då vi 
begränsat tillstånden i \eqref{idres}, är detta klart ekvivalent med att väntevärdet \eqref{GBcon} försvinner vid 
det semi-klassiska gränsvärdet för alla dessa koherenta ''mellanamplituder'' som förekommer då vi insätter resolutionen 
av identiteten för SU(2) i BF-teori med koherenta tillstånd. På detta vis kan vi sedan konstruera en spinnskum amplitud 
som sammanfaller med EPRL-modellens amplitud för Lorentzisk signatur. Det kan vara intressant att notera att dessa
amplituder inte sammanfaller för Euklidisk signatur då $\gamma >0$.     

\subsubsection{Den kosmologiska konstanten och divergenser}

Eftersom SKG är ultraviolett ändlig är det intressant att notera att spinnskum modellerna inte är det utan introduktionen av en kosmologisk konstant. 
Spinnskum modellerna innehåller två typers divergenser: bubblor (eng. \emph{bubble divergences}) och spikar (eng. \emph{spike divergences}) som är mycket starkt 
relaterade och kanske t.o.m. endast olika sidor av samma slant. Bubblorna uppstår 
då delar av trianguleringen som beskriver en hyperyta inne i 2-komplexet inte är begränsade, utan kan vara obegränsat stora. Spikarna i sin tur uppstår då 
kantlängderna i trianguleringen kan bli ''oändligt'' långa. I Ponzano-Regge modellen förväntar vi oss att spikarna är relaterade till diffeomorfism symmetri \cite{22, 23}. 
Men i \cite{24} har det också föreslagits att dessa divergenser i Ponzano-Regge modellen skulle relateras till summan över ''pariteter'' i teorin\footnote[1]{Se \cite{edcar} för en intressant diskussion om paritets- och tidsinversionstransformationerna för Holst aktionen och spinnskum modellerna.}. Nämligen godkänner 
Plebanski-formuleringen båda förtecknen för determinanten av metriken (jfr. \eqref{pleb}) och växlandet mellan dessa sektorer leder till divergenser. En annan relaterad 
kalkyl har gjorts i \cite{aldo30} där den mest divergerande kontributionen till EPRL modellen beräknats för själv-energin för spinnskummet. Själv-energin består 
av två 4-simplex där fyra av båda 4-simplexets kanter kopplas ihop så, att endast två kanter står för gränsdatan. Denna kalkyl visar att den mest divergenta kontributionen 
divergerar logaritmiskt och kan relateras till växelverkan mellan två kvanta av rummet, ett kvanta och ett antikvanta användande terminologin introducerad i \cite{aldo40}. 
Däremot har det visats att EPRL modellen med en kosmologisk konstant är fri från divergenser \cite{ha1, ha2, fr1, fr2}. Detta 
beror på att den kosmologiska konstanten implementeras via kvantdeformerade grupper, i vårt fall SL$_q(2, \mathbb{C})$ och transitionsamplituderna gäller 
spinnen $j_f < \Lambda$.  

\subsection{Semi-klassisk analys av EPRL modellen \label{semian}}

Den semi-klassiska analysen av spinnskum modellerna \cite{bar1, bar2} (se \cite{bar3} för en review) är viktig för oss, för att ha kännedom om modellernas geometriska 
egenskaper när modellen börjar närma sig den klassiska formuleringen. Eftersom hela spinnskum ideologin är 
baserad på diskreta strukturer, kan man endast vid detta semi-klassiska gränsvärde hoppas att nå en Regge-lik diskret 
formulering av den allmänna relativitetsteorin. Man brukar därför göra en skillnad mellan det semi-klassiska gränsvärdet av teorin, 
och gränsvärdet för kontinuum. Medan det fortfarande är oklart på vilket sätt SKG eller spinnskum modellerna överensstämmer med 
det senare gränsvärdet, rumtids formuleringen av allmän relativitetsteori, står resultatet av det semi-klassiska gränsvärdet för spinnskum modellerna, 
då man fixerar diskretiseringen, rätt klart. Det är Regge gravitation, åtminstone för Euklidisk signatur.

Man kan formulera det semi-klassiska gränsvärdet enligt följande. Vi inför en extra parameter $\lambda$ i gränsvärdet, m.h.a. 
vilket vi kan förstora $j$ och förminska $\hbar$ enligt $j \rightarrow \lambda j$ och $\hbar \rightarrow {\hbar\over \lambda}$, då $\lambda \gg 1$.  
Då vi väljer $\lambda$ stort kommer modellen att approximera en klassisk diskret geometri. En nyckel till detta resultat ligger i användningen 
av koherenta tillstånd, eftersom de har egenskapen att deras ''vågpaket'' är störst på klassiska extrinsiska och intrinsiska geometrier. Tanken är att
använda en diskret effektiv aktion given med koherenta tillstånd och sedan använda stationär-fas metoder för att analysera dess asymptotiska, semi-klassiska 
värde.

För att introducera denna metod kort, kan vi se på SU(2) BF-teori, eftersom vi redan introducerat koherenta tillstånd för SU(2) BF-teori. 
Dess 4-simplex amplitud (jfr. \eqref{bfpart}) med koherenta tillstånd (se föregående delsektion) kan skrivas som 
\be
A^{4-S}_{BF}(j, \mathbf{n}) =  \int \prod_{a=1}^5 dg_a \prod_{1\leq a \leq b \leq 5}^{5} \langle n_{ab} | g_a^{-1}g_b | n_{ba} \rangle^{2j_{ab}},
\ee 
där $j_{ab}$ är 10 spinn associerade med ytor och $n_{ab}$ dessa ytors normaler vilka är 20 st totalt (eftersom $n_{ab} \neq n_{ba}$) och vi använt 
egenskapen \eqref{koheg}. Vi kan skriva om denna amplitud som 
\begin{align}
A^{4-S}_{BF}(j, \mathbf{n}) &=  \int \prod_{a=1}^5 dg_a \prod_{1\leq a \leq b \leq 5}^{5} \exp{S_{j, \mathbf{n}}(g)} \\
S_{j, \mathbf{n}}(g) &= \sum_{a < b=1}^5 2j_{ab}\ln \langle n_{ab} | g_a^{-1}g_b | n_{ba} \rangle. \label{4saktion} 
\end{align}
\eqref{4saktion} är en aktion för ett 4-simplex, och som sådant kan vi studera dess asymptotiska beteende. För att använda 
stationära fasmetoder för dess analys, måste vi utöka lite på dessa metoder eftersom aktionen \eqref{4saktion} är komplex \cite{fri23, fri24}. Detta går 
enkelt genom att kräva att dess reella del är maximal vid det stationära värdet. Punkter som är både stationära och maximerar den 
reella delen av aktionen kallas \emph{kritiska}. Parametrarna $j_{ab}$ och $\mathbf{n}$ bestämmer den yttre gränsen på 
4-simplexet och därmed dess yttre gränsdata. Antalet kritiska punkter bestäms därför av den yttre gränsdatan och 
aktionens asymptotiska värde kommer därför att bestämmas av denna data. Denna 4-simplex amplitud har följande form i EPRL-modellen
\be
A^{4-S}_{EPRL} \sim {1\over \lambda^{12}}\Big[N_+ \exp\big(i \lambda \gamma \sum_{a<b} j_{ab}\Theta_{ab}\big) + N_- \exp\big(-i\lambda \gamma \sum_{a<b} j_{ab}\Theta_{ab} \big)  \Big],
\ee 
där $N_\pm$ är konstanter som inte ökar eller minskar med $\lambda$ och $\Theta_{ab}$ är underskottsvinkeln bekant från Regge gravitation. Amplituden har två termer p.g.a. att man kan associera ett paritetsrelaterat 4-simplex till varje lösning av stationär fas och maximal reell del av aktionen. 

Om detta resultat utökas till flere simplex kan vi, då $\gamma > 1$ och för Euklidisk signatur av FK modeller med 
Barbero-Immirzi parameter, konstatera att den asymptotiska analysen av aktionen i dessa spinnskum modeller delar sig i två sektorer, lösningar 
på rörelse-ekvationerna som är geometriska och icke-geometriska. De icke-geometriska lösningarna är exponentiellt dämpade då $\lambda j_f \gg 1$ trots att
vi fortfarande har två kontributioner till den asymptotiska amplituden, geometriska och icke-geometriska. Om man för hand begränsar sig till den geometriska sektorn 
får man amplituden
\be
W^{FK}_{\Delta^*}(j_f) \sim {c\over \lambda^{33n_e - 6n_v - 4n_f}} \exp\Big(i\lambda S_\mr{Regge}(j_f, \Delta^*)\Big), \label{asmpv}
\ee
där $n_e, n_v$ och $n_f$ är antalet kanter, spetsar och ytor i 2-komplexet $\Delta^*$. Det är värt att notera att aktionen \eqref{asmpv} inte längre beror av $\gamma$. Om
detta resultat kan generaliseras till en modell med flere 4-simplex återstår att se, men indikationer åt detta håll finns. Arbete gällande den asymptotiska analysen av både de Euklidiska \cite{ha11} och de Lorentziska \cite{ha12} modellerna med ett godtyckligt antal 4-simplex har visat att asymptotiken delar in sig i tre sektorer. Dessa sektorer skiljer sig i geometri. En sektor för icke-degenererad Lorentzisk diskret geometri, en för degenererad Lorentzisk geometri som kan beskrivas m.h.a. av icke-degenererad Euklidisk diskret geometri och en sektor för degenererad Lorentzisk geometri som kan beskrivas som en vektor geometri. Regge geometri kan också nås inom denna analys, om man delar in de kritiska konfigurationerna av den asymptotiska analysen i underkomplex (eng. \emph{subcomplex}) och limmar ihop dem genom att passligt betrakta deras orienterade 
4-simplex volym.

\subsubsection{Gravitationspropagatorn \label{grproa}}

Ett mycket intressant resultat för spinnskum modellerna kommer ifrån konstruktionen och beräkningen av gravitationspropagatorn för den Euklidiska 
EPRL modellen. Idéerna till denna konstruktion introducerades i \cite{grpro}. Dessa tillämpades på Barret-Crane modellen i \cite{wpro, pro2, pro3} och det konstaterades att 
Barret-Crane modellens spetsamplituds asymptotik inte överensstämde med Regge gravitation men i EPRL modellen fungerade asymptotiken \cite{euel, graap, hmm, hmm2}.  

Eftersom EPRL modellen är konstruerad som en diskretisering, kommer denna be\-räkning att resultera i gravitationspropagatorn i en 
diskretisation av den allmänna relativiteten, Regge modellen. Dessutom, som nämndes tidigare, kan denna beräkning endast göras för 
ett 4-simplex på spinnskummets yttre gräns och med Euklidisk signatur. Generaliseringen till ett större antal 4-simplex på den yttre gränsen 
har hittills endast gjorts inom en annan spinnskum modell, Barrett-Crane modellen, inte för EPRL modellen och generaliseringen till Lorentzisk signatur är 
för tillfället ett öppet problem. Trots detta, är denna beräkning en av de största succéer som EPRL modellen haft. Den visar att EPRL modellen
definitivt innehåller någonting korrekt gällande modellens gravitationsdynamik. Dessutom är en illustrering av denna beräkning mycket belysande gällande 
strukturen av spinnskum modellerna.  

Vi börjar med att definiera n-punkts funktionerna för kvantfältteori i stig-integral representationen i Euklidisk signatur som
\be
W(x_1,..., x_n) = {1\over Z}\int\mc{D}\phi\, \phi(x_1)\cdots \phi(x_n) e^{-S^E[\phi]}, \label{eunp}
\ee
där $\phi$ är kvantfältet, $x_1,...,x_n$ punkterna i bakgrunds rumtiden, $S^E[\phi]$ den Euklidiska aktionen och $Z = \int\mc{D}\phi\, e^{-S_E[\phi]}$. Vår uppgift är att
kunna skriva om detta uttryck så, att det passar ihop med tanken om en bakgrundsfri formulering av gravitation, grundstenen för SKG och spinnskum modellerna. Vi börjar 
med att dela in uttrycket $\eqref{eunp}$ i två delar, genom att dela in den totala rumtiden i två delar: en del $R$ och dess komplement $R^c$. Punkterna $x_1,..., x_n$
befinner sig på den yttre gränsen av $R$, vilken vi hädanefter kallar $\Sigma$ och $\vp$ är $\phi$:s restriktion till $\Sigma$. I detta fall kan vi skriva om \eqref{eunp} som
\begin{align}
W(x_1,..., x_n) &= {1\over Z}\int\mc{D}\vp\, \vp(x_1)\cdots \vp(x_n) W_R[\vp, \Sigma] W_{R^c}[\vp, \Sigma], \label{eunp1} \\
W_R[\vp, \Sigma] &=  \int_{\phi|_\Sigma = \vp}\mc{D}\phi_R\, e^{-S^E_R[\phi_R]}, \label{Wth}
\end{align}
där $S^E_R$ är aktionens restriktion till regionen $R$, och integralen \eqref{Wth} går över fälten inom $R$ som begränsas av fälten $\vp$ på dess 
yttre gräns $\Sigma$. $W_{R^c}[vp, \Sigma]$ analog till \eqref{Wth}, men den gäller fälten begränsade till R:s komplement $R^c$. Med denna omskrivning av 
uttrycket \eqref{eunp} som bättre passar spinnskum formalismen, kan vi tänka oss att funktionalen $W_{R^c}[\vp, \Sigma]$ i en växelverkande teori i en passlig 
approximation närmar sig funktionalen för en fri teori och blir till sin form Gaussisk, medan själva växelverkan kan begränsas att gälla endast regionen $R$. I detta 
fall kan vi välja $W_{R^c}[\vp, \Sigma]$ som om det gäller en fri teori enligt
\be
W^0_{R^c}[\vp, \Sigma] =  \int_{\phi|_\Sigma = \vp}\mc{D}\phi_{R^c}\, e^{-S^{0}_{R^c}[\phi]} = \Psi_{\Sigma}[\vp]. 
\ee 
I ovanstående uttryck står $S^0_{R^c}[\phi]$ för den Euklidiska aktionens fria del i regionen $R^c$. Integralen är Gaussisk och bestämmer 
tillståndet på den yttre gränsen av $R$, som vi kallar $\Psi_{\Sigma}[\vp]$. 

Efter detta lönar det sig att komma ihåg att vi arbetar med en bakgrundsfri teori och är intresserade av n-punkts funktionerna för gravitation. I detta specifika fall, betyder det att
$\phi$ är gravitationsfältet och måttet i integralen är bakgrundsinvariant. Av detta följer att \eqref{Wth} propagatorn inte beror av lokala deformationer 
av $\Sigma$ och därför är \eqref{Wth} inte beroende av $\Sigma$. En annan följd är att geometrin av $\Sigma$ inte bestäms av någon bakgrunds geometri (eftersom ingen 
finns) men av gravitationsfältet på $\Sigma$. M.h.a. denna heuristik, kan vi då skriva om uttrycket  \eqref{eunp} i en form som kan tillämpas av 
spinnskum modellerna:
\be
W(x_1,..., x_n) = \langle W|\vp(x_1)\cdots \vp(x_n)|\Psi_q \rangle={1\over Z}\int\mc{D}\vp\, \vp(x_1)\cdots \vp(x_n) W[\vp] \Psi_q[\vp],  \label{eunpf}
\ee 
där $\Psi_q[\vp]$ är ett tillstånd som är som störst kring en klassisk 3-geometri, $q$. Eftersom detta uttryck är kovariant, måste vi ännu klargöra 
betydelsen av koordinaterna. Dessa kan helt enkelt tänkas som definierade i avseende å geometrin $q$. T.ex. för 4-punkts funktionen kan vi ta $t_1 = t_2 = 0$ och 
$t_3 = t_4 = T$ (vi använder beteckningen $x = (t, \vec{x}$), och välja $q$ som geometrin för en rektangulär låda med höjden 
$T$ och bredden $L$ och $x_i$ är egendistanser från lådans kanter.  

Vi kan nu använda denna idé för att formulera en beräkning av gravitationspropagatorn eller två-punktsfunktionen för den Euklidiska EPRL-modellen. Som 
mångfald $R$ väljer vi 4-kulan och dess yttre gräns $\Sigma$ har därför topologin av en av en 3-sfär, $S^3$. Med denna yttre gräns associerar vi ett Hilbertrum
$\mc{H}_{\Sigma}$ som består av spinn-nätverk. Sedan väljer vi ett Euklidiskt 4-simplex som geometri för den yttre gränsen. Detta innebär att vi diskretiserar rummet
$\Sigma$ i fem tetraedra som är kombinerade så att de bildar ett 4-simplex. Denna konstruktion av gränsdata gör att det semiklassiska tillståndet $|\Psi_q\rangle$ 
är som störst på en intrinsisk och extrinsisk geometri $q$ som är platt. 

Efter detta steg måste vi ännu ange $\langle W|$ i \eqref{eunpf}. Denna kan vi ta från en expansion enligt en GruppFältTeori (GFT) (eng. \emph{group field theory}) formulering av 
EPRL-modellen \cite{eprlgft1} (se delsektion \ref{avslut} och t.ex. \cite{friGFT, eprlgft2} för en introduktion). I denna formulering består spinnskum modellerna av expansioner i en parameter och vi kan beräkna två-punkts funktionen inom denna formulering till första ordning i en spets-expansion. M.a.o. beräknas gravitationspropagatorn inom approximationen där spinnskummet innehåller en spets vilken resulterar i ett 4-simplex spinn-nätverk på spinnskummets yttre gräns. 

Till slut beräknar vi två-punktsfunktionen, där den metriska operatorn anges av $q^{ab} = \delta^{ij}E^a_i E^b_j$ och med ovan angivna data, från
\begin{align}
G^{abcd}(x, y) = \langle q^{ab}(x) q^{cd}(y) \rangle - \langle q^{ab}\rangle \langle q^{cd}\rangle \label{2pf} \\
\end{align}    
inom approximationen $j \rightarrow \infty, \gamma \rightarrow 0, j \gamma = konst.$. Detta ger oss precis två-punkts\-funktionen för 
Regge gravitation inom samma approximation \cite{euel}. Orsaken att gränsvärdet $\gamma \rightarrow 0$ måste tas, beror på att spinnskum modellerna
tillåter icke-geometriska konfigurationer och dessa existerar inte inom diskret Regge gravitation. Varför de kontrolleras av Barbero-Immirzi parametern
står för tillfället oklart. Generaliseringen av detta resultat till flere spetsar i spinnskummet  eller Lorentzisk signatur står också oklart. Däremot har t.ex. 
den gravitationella tre-punktsfunktionen beräknats \cite{mingY}. 

\subsection{Spinnskum modellernas status just nu \label{avslut}}

Spinnskum modellerna ger oss en stig integral av den gravitationella växelverkan som en summa av 2-komplex med produkter av amplituder på 
2-komplexets byggstenar. Teorin är genomgående lokalt Lorentz kovariant (se t.ex \cite{sica}), vilket är mycket intressant eftersom Planck-längden är det kortaste avståndet som modellen accepterar. Vi har sett att man inom denna formulering semi-klassiskt kan beräkna gravitationspropagatorn (delsektion \ref{grproa}), vilket bevisar att modellen innehåller åtminstone en del av den korrekta gravitationsdynamiken, dynamiken som SKG inte till dags dato klarat av att beskriva. Modellen har visats vara 
starkt relaterad till SKG \cite{dy1, dy2, wolf1, wolf2}, man kan också addera materia till den \cite{mat1, mat2} och man kan skapa en intressant parametrisering av 
den m.h.a. twistorer \cite{twi1, twi2, twi3, twi4, twi5} (eng. \emph{twistors}). 

Kosmologi har undersökts inom den enklaste approximationen av modellen, en transition mellan två dipol\footnote[1]{En dipol 
graf består av två noder och fyra stigar som sammanlänkar dem.} grafer som spinn-nätverk på den yttre gränsen och med en spets mellan dem i spinnskummet \cite{ksmo}.
Denna modell beskriver en transition mellan två koherenta tillstånd som har sina största värden på homogena och isotropa metriker. Man kan också inkludera den kosmologiska
konstanten i denna approximation \cite{ksmkon} och tankar om hur man kunde börja betrakta inhomogena modeller har presenterats i \cite{inho11}. 
   
Spinnskum modellerna har m.a.o. visat lovande resultat och en positiv utveckling sedan deras födsel, men som alltid finns det obesvarade frågor. En av dessa gäller 
måttet på stig integralen. Då man ålägger begränsningarna på stig integralen kommer dessa, förutom att begränsa oss till gravitation, också att modifiera måttet på 
integralen \cite{alx}. Denna modifikation är inte problematisk, men eftersom teorin skall vara bakgrundsfri, sätter detta restriktioner på hur måttet får se ut. Detta leder till 
en tvetydighet i teorin. BF-måttet som använts i denna review är ett mått som är kompatibelt med bakgrundsfrihet, men det finns andra. Se t.ex. måttet i \cite{ero} för ett alternativ 
som också är kompatibelt med bakgrundsfrihet och som motiveras från SKG.   

En annan fråga gäller projektorn i EPRL modellen. Man kan skriva partitionsfunktionen för EPRL modellen m.h.a. en projektor så som i BF-teori (se \eqref{zbf4}). 
Orsaken till detta är definitionen på den formellt definierade idéen bakom spinn-nätverken som en projektor på lösningarna av Hamiltonbegränsningen (se delsektion \ref{ide}). 
Inom EPRL modellen kommer den sk. projektorn inte att fungera som en projektor eftersom $P_{EPRL}^2 \neq P_{EPRL}$. Detta är problematiskt eftersom det gör 
att den nuvarande bästa modellen för en spinnskum modell inte överensstämmer med de gamla konsistenskraven på en spinnskum modell \cite{baez232}. Dessa konsistenskrav 
är en konsekvens av att man kräver att modellen inte skall bero av trianguleringen av mångfaldet så att man kan passa in flere olika spinnskum på samma triangulering vilka 
alla beskriver samma fysikaliska process, bakgrundsfriheten. Att EPRL projektorn inte är en projektor motsäger detta krav. Situationen är trots detta inte katastrofal eftersom den
Euklidiska EPRL modellen kan modifieras så, att den består av en projektor som satisfierar ${P^e_{EPRL}}^2 = P^e_{EPRL}$ \cite{rpe1, rpe2}. I den Lorentziska modellen 
är denna fråga ännu öppen.

En tredje utveckling som är relevant för spinnskum modellerna, är GruppFältTeorierna (GFT). Inom dessa modeller skapas spinnskummet av en expansion av spetsamplituder
i en parameter $\lambda$. Expansionen är analog med expansionen i Feynman diagram inom kvantfältteori \cite{gftr, gftrr}. Propagatorn som man anger rörelsemängds 4-vektorn på, är analog med spinn på ytorna i spinnskummet och konservering av rörelsemängden i spetsarna i Feynmandiagram, representeras av konservering av spinn i spetsarna som implementeras av sammanflätarna i spetsarna på spinnskummet. Spetsamplituden representerar växelverkan. Meriten av denna formulering är att man inte längre behöver fundera över spinnskummets beroende av diskretisationen och diffeomorfismerna inkluderas automatiskt. Dessa två frågor som är svåra inom den vanliga formuleringen av spinnskum modellerna är alltså lättare inom GFT, men å andra sidan blir svårigheten inom denna formulering den fysikaliska rollen av $\lambda$ och konvergensen av 
expansionen. Den enklaste fysikaliska tolkningen av GFT är att den representerar 3-dimensionell kvantgravitation med en dynamisk topologi, men den fysikaliska tolkningen as dessa modeller är ännu en öppen fråga.

\large
\begin{center}
\textbf{Tack}
\end{center}
\normalsize

Jag vill tacka hela kvantgravitationsgruppen i CPT Marseille för ett njutbart år av forskning och en mycket öppen och professionell forskningsatmosfär. Jag vill speciellt 
rikta mitt tack till Marios Christodoulou, Muxin Han, Aldo Riello, Carlo Rovelli, Christian Röken, Simone Speziale och Wolfgang Wieland, utan vilka mina första relativt kaotiska bekantskaper med spinnskum modellerna antagligen endast skulle ha förblivit en röra.

 \end{document}